\documentclass[%
 reprint,
superscriptaddress,
 amsmath,amssymb, 
 aps,
prb,
floatfix,
]{revtex4-2}

\usepackage[normalem]{ulem}
\usepackage{xcolor}
\usepackage{graphicx}% Include figure files
\usepackage{dcolumn}% Align table columns on decimal point
\usepackage{bm}% bold math
\usepackage{hyperref}% add hypertext capabilities
\usepackage{braket}

\begin{document}

\preprint{APS/123-QED}

\title{Entanglement Action for the Real-Space Entanglement Spectra of Composite Fermion Wave Functions}

\author{Greg J. Henderson}
\affiliation{Rudolf Peierls Centre for Theoretical Physics, Parks Road, Oxford OX1 3PU, United Kingdom}
\author{G J Sreejith}
\affiliation{%
 Indian Institute of Science Education and Research, Pune 411008, India
}
\author{Steven H. Simon}
\affiliation{Rudolf Peierls Centre for Theoretical Physics, Parks Road, Oxford OX1 3PU, United Kingdom}

\date{\today}% It is always \today, today,

\begin{abstract}
 We argue and numerically substantiate that the real-space entanglement spectrum (RSES) of composite fermion quantum Hall states is given by the spectrum of a local boundary perturbation of a $(1+1)$d conformal field theory (CFT), which describes an effective edge dynamics along the real-space cut. The cut-and-glue approach suggests that the low-lying RSES is equivalent to the low-lying modes of some effective edge action. The general structure of this action is deduced by mapping to a boundary critical problem, generalizing work of Dubail, Read, and Rezayi [PRB 85, 11531 (2012)]. Using trial wave functions we numerically test our model of the RSES for the $\nu=2/3$ bosonic composite fermion state.
\end{abstract}

\maketitle

%\tableofcontents

\section{Introduction} \label{Intro}
Ever since its discovery, the fractional quantum Hall effect has challenged several preexisting paradigms in condensed matter physics. The various filling fractions correspond to different phases of matter which cannot be distinguished by the traditional methods of symmetry breaking and order parameters, and are examples of \textit{topological phases of matter}\cite{Nayak, Wen2016, Duncan2017}. New methods have been required to characterize such phases of matter.

One such method that has proved particularly useful is to bipartition the system and study the \textit{quantum entanglement} between the two subsystems when the whole system is in its ground state. Perhaps the simplest measure of the entanglement between the two subsystems is the so-called ``entanglement entropy''. As argued separately by Kitaev and Preskill \cite{Kitaev2006}, and Levin and Wen \cite{Levin2006}, if one bipartitions a topological phase in real-space then one could extract the total quantum dimension of the underlying topological quantum field theory from the entanglement entropy.

Entanglement entropy is, however, not the full story. Given the bipartition into subsystems, $A$ and $B$, then all the information regarding entanglement between the two pieces is contained in the reduced density matrix of $A$, $\rho_A$, or equivalently $\rho_B$. In this paper, we will be interested in the \textit{entanglement spectrum} (ES), which is the spectrum of eigenvalues of $- \ln \rho_A$. When the bipartition is in real-space (i.e. when we partition the system into two spatial regions) then the ES is termed the \textit{real-space entanglement spectrum} (RSES).

As proposed and numerically substantiated by Li and Haldane, the ES may contain certain universal features related to the topological phase beyond the total quantum dimension \cite{Li2008}. In particular, they found that the low-lying ES of an orbital partition, which approximates the real-space partition, of the Moore-Read trial wave function and the corresponding coulomb ground state resembled that of the (perturbed) CFT which describes the edge dynamics of the system.

Since Li and Haldane's discovery, many other works have shown that these features extend to other fractional quantum Hall states \cite{Sterdyniak2012, Rodriguez2012, Thomale2010, Dubail2012a, Regnault, Lauchli2010, Zhang2012, Hermanns2011, Papic2011, Schliemann2011, Sterdyniak2011a, Yan2019}, integer quantum Hall states \cite{Rodriguez2009, Turner2010} and general topological phases of matter \cite{Fidkowski2010, Regnault2011, Prodan2010, Pollmann2010, Hughes2011, Alexandradinata2011, Yao2010, Dubail2011}. Further, some general arguments have been given as to why such features are apparent \cite{Chandran2011, Qi2012, Dubail2012, Swingle2012}. Most relevant to this paper are the works of Qi, Katsura and Ludwig (QKL) \cite{Qi2012}, and Dubail, Read and Rezayi (DRR) \cite{Dubail2012}. In the work of QKL, it is suggested that we can understand the RSES by modelling the interaction between the two real-space regions as only occurring between the edge degrees of freedom. Thus, one can model the RSES as the ES of two interacting edges with opposite chirality. This has now been termed the \textit{cut and glue} approach. By an intrinsically different approach, DRR rigorously showed, under one assumption, that the RSES of certain quantum Hall trial wave functions is the same as the spectrum of an operator, known as the entanglement action, which is an integral along the real-space cut of a sum of local operators that belong to a CFT defined on the real-space cut. DRR termed this the ``scaling property'' of the RSES. The trial wavefunctions that DRR examined were those that can be expressed as conformal blocks of a CFT. This includes, for example, the Laughlin\cite{Laughlin1983}, Moore-Read\cite{Moore1991} and the Read-Rezayi\cite{Read1999} series wave functions. In each case the CFT used to express the wave function is the same CFT used to describe the RSES. Interestingly, the work of DRR allowed them to understand the RSES of large but \textit{finite} system sizes. By renormalization group (RG) arguments, they were able to determine how the coefficients of the various operators, comprising the entanglement action, scaled with the system size. Thus, at large but finite system sizes one would only need the most relevant terms to accurately reproduce the RSES.

Not all trial wave functions, however, are covered by the arguments of DRR. Many experimentally prominent filling fractions have been described very well by the \textit{composite fermion} (CF) trial wave functions \cite{Jain1989, Jain2007}. These wave functions cannot be expressed directly as conformal blocks, but instead can be expressed as appropriately symmetrized conformal blocks \cite{Hansson, Kvorning2013}. This symmetrization procedure means one cannot directly extend the arguments of DRR to the CF wave functions. Generally speaking, the intricate structure of the trial wave functions considered by DRR has allowed the physics of such states to be well understood, whereas it is still not known if much of the same physics still applies to the CF wave functions \cite{Simon2020}.

There is comparatively little literature on RSES of CF wave functions. Rodriguez et al. \cite{Rodriguez2013} showed how to numerically calculate the RSES for CFs and Davenport et al.\cite{Davenport2015} suggested it could be modelled by the spectrum of a Hamiltonian of weakly interacting fermions. In the work of Davenport et al., the fermionic model was only shown to work over a subset of the RSES with a fixed number of particles in each subsystem. It is also not clear how the Davenport et al. model is connected with the previous works on the RSES, where one would expect the RSES to be described by an edge theory of the CF state, which would not be equivalent to a theory of weakly interacting fermions (i.e. the weakly interacting fermion theory would correspond with an integer quantum Hall edge). Further, using the methods described in Refs. \cite{Lundgren2013, Chen2013}, Cano et al.\cite{Cano2015} showed what form the RSES should take in the thermodynamic limit for all chiral abelian quantum Hall states, of which the composite fermion states are an example, by applying the cut and glue method and directly modelling the interactions between the edges. However, it is not obvious from their approach what form the finite size corrections take, nor has their result been numerically validated yet.

In the current paper we argue, and numerically substantiate, that starting from the QKL cut and glue approach and applying boundary CFT techniques similar to DRR, the RSES of composite fermion wave functions can be understood as the spectrum of an entanglement action operator that is a sum of integrals along the real-space cut of local operators belonging to a CFT on the real-space cut. The CFT in question, by assumption, is the same as that which can describe the \textit{minimal} edge of the corresponding CF state. In short, we wish to argue that the DRR \textit{scaling property} holds for the CF wave functions.

In Sec.~\ref{SurfaceCriticalTheory} we argue that the DRR scaling property holds for the RSES of chiral abelian quantum Hall states, starting from the QKL cut and glue approach. Much of the discussion of Sec. \ref{SurfaceCriticalTheory} will have a large overlap with the work of DRR and QKL. However, the methods of both of these references need to be combined to argue the existence of the scaling property. Then in Sec. \ref{ConstructionOf23Model}, we use the result of Sec. \ref{SurfaceCriticalTheory} to develop an analytic model of the RSES entanglement spectrum of the bosonic $\nu = 2/3$ composite fermion wave function on the sphere with real-space cut along the equator. We expand this model order by order in the inverse system size and consider truncating the expansion either at the lowest nontrivial order or next the lowest order. Finally, in Sec. \ref{Section:NumericalTests}, we present two numerical tests of these models: one showing that the models can be fitted well to the numerically calculated RSES, over multiple sectors with varying number of particles in each subsystem, and another demonstrating that the parameters of the models scale with the system size as predicted by the renormalization group arguments of Sec. \ref{SurfaceCriticalTheory}. 

\section{Entanglement Spectrum as a Surface Critical Problem} \label{SurfaceCriticalTheory}
We now wish to argue that the DRR scaling property holds for the RSES of CF wave functions. The argument we present here covers the chiral abelian states, which includes the principal Jain sequence, and can be generalised to many other quantum Hall states. 

This argument is summarised as follows. First, we review the cut and glue approach, which then allows us to state our starting assumption. In short, this assumption is that the RSES can be understood as the ES between coupled chiral and anti-chiral quantum Hall edges in their ground state. We will assume that the edges are minimal (i.e., not reconstructed, for example) and we will outline the structure of such edge theories. To compute the reduced density matrix of either edge, one must first understand the structure of overlaps with the edge ground-state. We demonstrate that the problem of computing overlaps with the edge ground state can be converted to a problem of computing the correlators of a CFT on the semi-infinite cylinder with a \textit{local boundary perturbation}. The CFT is that which describes the two uncoupled edges and the boundary perturbation encodes all information related to the ground state of the coupled edges. As we will be interested in overlaps that correspond to modes of the correlators with wave lengths comparable to the circumference of the cylinder, we then analyze the case of a large real-space cut through renormalization group (RG) arguments (for a discussion of RG in the presence of boundaries we refer the reader to chapter 7 of Ref. \cite{Cardy2015}). Loosely speaking, as we increase the real-space cut length the effective boundary perturbation changes according to an RG transformation, where it will flow to some \textit{fixed point boundary condition}, which will be detailed later. Hence, for large real space cuts the boundary perturbation is equivalent to this fixed point boundary condition with \textit{irrelevant} perturbations. From our understanding of these overlaps we then simply compute the reduced density matrix of one of the edges. The final result being that the reduced density matrix takes the form $\rho = e^{-S_{ES}}$. $S_{ES}$ is expressed in radial quantisation as, $S_{ES} = \sum_i \frac{\alpha_i}{L^{h_i - 1}} \oint \frac{dz}{2 \pi i} z^{d_i - 1} \phi_i(z)$, where $\phi_i(z)$ are a complete set of operators for our single edge, $h_i$ are the corresponding scaling dimensions, $L$ is the real space cut length and $\alpha_i$ are numerical constants to be determined. We can then see that for large $L$ we only need to keep terms with a low scaling dimension to model the RSES.

The boundary CFT methods we use are identical to those used by DRR and are now common in the boundary CFT literature as a way of classifying the massive infra-red fixed points of two dimensional field theories \cite{Cardy2017, Andrei2020, Konechny2017}, and have been used to study entanglement in CFTs and Chern-Simon's theories \cite{Das2015, Wong2018}. Our novel contribution is to combine these methods with the QKL cut and glue approach and to identify the appropriate boundary conditions for the case of the CF states.

QKL also use certain boundary CFT techniques and the methods of Calabrese and Cardy \cite{Calabrese2006, Calabrese} to obtain an expression for the reduced density matrix in the large real-space cut limit. However, here will not apply the methods of Calabrese and Cardy, as they do not allow us to understand the finite size corrections and it is not clear how they should be applied in the presence of multiple edge modes.

Before moving on the reader should note two caveats. First, in applying the cut and glue method to composite fermion states we have implicitly assumed that each CF wave function is the exact ground state of some reasonably physical \textit{gapped} Hamiltonian, often known as a parent Hamiltonian. One can always construct a local Hamiltonian for which a given wave function is the ground state, however there is no guarantee in general that it is gapped \cite{Greiter2018, Chertkov2018, Qi2019, Cubitt2015}. Finally, the application of the cut and glue method only models the ES below the \textit{entanglement gap}, \cite{Li2008, Regnault, Thomale2010} i.e., for a range of low  entanglement energies as we shall discuss in a moment.

\subsection{Cut and glue approach}\label{sec:cutGlue}
The central idea behind the cut and glue approach is that if the ground state only has short range correlations, then under a spatial bipartition the most entangled states must be the edge states of our two spatial regions. Here we shall give a precise statement of this that will allow us to model the RSES of chiral abelian quantum Hall states.

Consider a system in a fractional quantum Hall ground state and a spatial bipartition into two regions $A$ and $B$. We can express the ground state, $\ket{G}$, as a Schmidt decomposition, $\ket{G} = \sum_i e^{-\xi_i /2} \ket{i_A}\otimes\ket{i_B}$, where $\xi_i$ are real numbers, and $\{ \ket{i_A} \}$ and $\{ \ket{i_B} \}$ form orthonormal bases for subsystems $A$ and $B$ respectively. From the definition given in the introduction, we see that the set $\{\xi_i \}$ is the real space entanglement spectrum. Several works have pointed out that there is a gap in the entanglement spectrum \cite{Li2008, Regnault, Thomale2010}. Let $\Xi$ be the set of indices, $i$, such that $\xi_i$ is below the entanglement gap. We will assume that for $i \in \Xi$, $\ket{i_A}$ and $\ket{i_B}$ are edge states of the $A$ and $B$ systems respectively, i.e., these kets do not have excitations deep in the bulk. We will further assume that $\ket{G_{\text{edge}}} = \sum_{i\in \Xi} e^{-\xi_i /2} \ket{i_A}\otimes\ket{i_B}$ is the ground state of some ``physical'' \textit{gapped} Hamiltonian only involving edges of the $A$ and $B$ subsystems, $H_{\text{edge}}$.

One can give an intuitively physical, but by no means rigorous, justification for this. Write the Hamiltonian of the system as $H = H_A + H_B + H_{\text{AB}}$, where $H_A$ and $H_B$ are the Hamiltonians for subsystems $A$ and $B$ respectively, and $H_{AB}$ is the interaction between them. If we turn off this interaction then one would expect at low energies the Hilbert space ``looks'' like $(\mathcal{H}_{A,\text{bulk}}\otimes\mathcal{H}_{A,\text{edge}})\otimes (\mathcal{H}_{B,\text{bulk}}\otimes\mathcal{H}_{B,\text{edge}})$. Further, one would also expect that at low energies of the full Hamiltonian $H$ we are still at low energies in the non-interacting Hamiltonian, $H_A + H_B$. Thus, we write $H = H_{\text{edge}} + H_{A, \text{bulk}} + H_{B, \text{bulk}} + \Delta H$, where $H_{\text{edge}}$ only involves the edge degrees of freedom (including the interactions between the edges), $H_{A, \text{bulk}}$ and $H_{B, \text{bulk}}$ are the Hamiltonians of the bulks of $A$ and $B$ repsectively, and $\Delta H$ includes the bulk-edge interaction, both between A and B and within either system, and the bulk-bulk interactions between the $A$ and $B$ systems. Even for the long range coulomb interaction, one would expect the energy scale for $H_{\text{edge}} + H_{A, \text{bulk}} + H_{B, \text{bulk}}$ to be larger than the energy scale for $\Delta H$. Thus, we expect $\Delta H$ can be considered a \textit{perturbation} of $H_{\text{edge}} + H_{A, \text{bulk}} + H_{B, \text{bulk}}$. Further, the ground state of $H_{\text{edge}} + H_{A, \text{bulk}} + H_{B, \text{bulk}}$ is such that both the $A$ and $B$ are in their individual ground states and the coupled edges are in the ground state of $H_{\text{edge}}$. This would then imply the assumption given in the previous paragraph.

As given here, this assumption is stronger than that given by QKL. There is now at least some numerical evidence that one can make such an assertion \cite{Yan2019, Sterdyniak2011a}. The main point is that this approach reduces the problem to finding the entanglement spectrum between the edges of $A$ and $B$ in $\ket{G_{\text{edge}}}$.

\subsection{Minimal edges of chiral abelian quantum Hall phases} \label{Section:MinEdgeGeneral}
Before going on to understand the entanglement structure of $\ket{G_{\text{edge}}}$, we must first detail the edge theory of these chiral abelian phases. In the following, we will assume that the required edge theory to describe the edges of $A$ and $B$ is that of a minimal edge. The Hamiltonian of our two interacting edges can be written, $H_{\text{edge}} = H_0 + perturbations$, where $H_0$ has full conformal symmetry and does not contain any interactions between the edges. The perturbations include all interactions between the $A$ and $B$ edges and any other terms involving only one edge that also break conformal symmetry. We shall now detail the structure of the edge $A$ under $H_0$, where the structure of the edge of $B$ is simply the anti-chiral copy. Here we shall only detail the aspects of the edge required for the following work. For fuller accounts on quantum Hall edges we refer the reader to Refs. \cite{Wen1995, Cano2014, Hansson2017, Chang} and for general discussion of conformal field theories see Refs. \cite{Ginsparg1988, Cardy2008, DiFrancesco1997}.

In the case of an Abelian quantum Hall phase, which is all we shall consider in this work, the (conformal) edge theory of $A$ in general will contain $p$ free chiral boson fields, denoted by $\varphi^{(i)}(x)$. Under the hierarchy construction, $p$ is simply the level of the hierarchy to which this state belongs. Furthermore, the boson fields are compactified, such that the vector of $U(1)$ charges of the fields, $a = (U(1) \text{ charge of } \varphi^{(1)}, U(1) \text{ charge of } \varphi^{(2)}, \dots)$, must belong to some integer lattice, $\Gamma$. We then define for any $a \in \Gamma$, a corresponding vertex operator $V_a(x) = :e^{i\sum_j a_j \varphi^{(j)}(x)}:$. All such vertex operators create an \textit{integer} multiple of electron\footnote{Here we use the word ``electron'' in a loose sense to refer to whatever particle the particular quantum Hall liquid is composed of.} charge at the edge; their exists a charge vector $t$ which belongs to the dual lattice, $t \in \Gamma^*$, such that $t \cdot a$ is the amount of electron charge created by $V_a(x)$. We can Fourier expand the vertex operators $V_a(x) = (2\pi / L)^{h_a} \sum_{m \in \mathbb{Z} - h_a} e^{-i\frac{2\pi m}{L}x} V_{a,m}$, where $h_a$ is the scaling dimension of this operator and $L$ is the length of the real-space cut\footnote{For simplicity, we have not written the time dependence of the edge operators.}. Further, it can be shown that $(V_{a,m})^\dagger = V_{-a, -m}$.

Denote a particular basis of $\Gamma$ by $\epsilon^i$. Through repeated operator product expansions between the set of vertex operators, $V_{\epsilon^1}, V_{\epsilon^2}, \dots, V_{\epsilon^p}, V_{-\epsilon^1}, \dots, V_{-\epsilon^p}$, one can generate \textit{all} local operators of the theory (see Appendix \ref{Appendix:BoundaryState} for details). Consequently, the Hilbert space of this edge theory forms an \textit{irreducible} representation of the chiral algebra, $\mathcal{A}$, generated by repeated operator product expansions between $V_{\epsilon^1}, V_{\epsilon^2}, \dots, V_{\epsilon^p}, V_{-\epsilon^1}, \dots, V_{-\epsilon^p}$ (where we use the term ``chiral algebra'' as defined in Ref. \cite{Moore1989}). Or in other words, all edge states can be generated by applying the modes of these vertex operators on the vacuum state, $\ket{0}$. 

For any local operator of the $A$ edge, $O(x)$, we will denote the corresponding anti-chiral version on the $B$ edge by $\overline{O}(x)$. An important point, that will be used later, is that if $V_a(x)$ creates some $q$ amount of a particular $U(1)$ charge on the $A$ edge, then $\overline{V}_a(x)$ will create some $-q$ of the same $U(1)$ charge on the $B$ edge. In other words, if $V_a(x)$ creates a particular excitation on the $A$ edge then $\overline{V}_{-a}(x)$ will create the \textit{same} excitation on the $B$ edge. This follows from the fact that the anti-chiral version of the $A$ edge is also the time reversed version. 

\subsection{Overlaps with the edge ground state}\label{sectionOverlaps}
In Sec. \ref{sec:cutGlue} we discussed how the problem of computing the RSES can be converted to that of computing the ES between the edges of $A$ and $B$ in $\ket{G_{\text{edge}}}$. We can understand this ES by first understanding the structure of overlaps with the edge ground state of the form, $\bra{a}\overline{\bra{b}}\ket{G_{\text{edge}}}$, where $\ket{a}$ and $\overline{\ket{b}}$ are edge states of $A$ and $B$ respectively.

As all states in this two edge system can be generated by modes of the various local fields, determining the above overlap is equivalent to the computation of correlators of the form, $\bra{0}\phi_1(x_1,\tau_1)\phi_2(x_2, \tau_2)\dots \ket{G_{\text{edge}}}$, where $\phi_i(x_i, \tau_i)$ is a local operator evolving in imaginary time, $\tau$, by $H_0$, $\tau_1 > \tau_2 > \cdots > 0$, and $\ket{0}$ is the vacuum of $H_0$. Furthermore, as we can write $\ket{G_{\text{edge}}} = \lim_{\tau \rightarrow \infty} e^{-\tau H_{\text{edge}}} \ket{0}/\braket{G_{\text{edge}} | 0}$, this is equivalent to computing correlators of fields, at $\tau > 0$, in a Euclidean quantum field theory, which is a CFT evolving according to $H_0$ for $\tau > 0$ and for $\tau \leq 0$ is a perturbed CFT evolving according to $H_{\text{edge}}$.

As any excitations along the real-space cut must be gapped, $H_{\text{edge}}$ must be gapped. Thus, for $\tau < 0$ we have short ranged correlations, with a correlation length related to the energy gap. Hence, if we ``integrate out'' the degrees of freedom for $\tau < 0$ we will be left with a CFT on a semi-infinite cylinder with a $\textit{local}$ boundary perturbation at the boundary, $\tau = 0$. In full, this means the action takes the form $S = S_{\textit{CFT}} + S_b$, where $S_b$ is local to the boundary and takes the form, $S_b = \sum_i \lambda_i \int_{\tau = 0} dx \phi_i(x)$, with $\phi_i(x)$ being local boundary operators and $\lambda_i$ being numerical coefficients that do not depend on the real-space cut length $L$ (when the real-space cut length is large compared to the correlation length). 

In what follows we shall only be interested in overlaps which correspond to modes of the correlators with wavelength comparable to $L$. Thus, we can use an RG procedure to compare the overlaps for two different real-space cut lengths. For now let us concentrate on real space cut lengths larger than some $L'$, $L>L'$. We can then take a particular real-space cut and perform standard RG until the circumference of the semi-infinite cylinder is $L'$. Under this RG the bulk action $S_{CFT}$ will be invariant, but $S_b$ will change. Thus, we can compare two real-space cuts by comparing the resulting $S_b$'s after this RG procedure. Let $S_b(L)$ denote the resulting boundary action \textit{after} this procedure.

For a large $L$, $S_b(L)$ will be very close to some fixed point which will enforce some \textit{fixed point boundary condition}. This boundary condition can be understood as a boundary state $\ket{G_*}$, where the correlators with this boundary condition are computed as $\bra{0} \phi_1(x_1, \tau_1) \phi_2(x_2, \tau_2)\dots \ket{G_*}$. One can think of $\ket{G_*}$ as the RG fixed point of $\ket{G_{\text{edge}}}$. We now wish to determine $\ket{G_*}$. 

Vertex operators create excitations on the edge which correspond to some bulk excitation that has been moved to the edge. Hence, a natural fixed point boundary condition that one would expect is that when we make a particular excitation on the $A$ edge this should be equivalent to making the corresponding excitation on the $B$ edge, as the two edges are not spatially separated. This then gives $V_{\epsilon^i}(x)\ket{G_*} = \overline{V}_{-\epsilon^i}(x)\ket{G_*}$. Written in terms of modes this gives,
\begin{equation} \label{boundaryCon}
    [V_{\epsilon^i,n} - \overline{V}_{-\epsilon^i, -n}]\ket{G_*} = 0
\end{equation}
This boundary condition is in fact the generalised form found by DRR. However, as we have applied the cut and glue approach, this boundary condition now has a physical interpretation. One can show by Schur's lemma that $\ket{G_*}$ is completely determined by Equation \ref{boundaryCon} up to a multiplication by a complex number (see Appendix \ref{Appendix:BoundaryState}). Furthermore, $\ket{G_*}$ has the property that, with a particular normalisation, $\bra{a} \overline{\bra{b}} \ket{G_*} = \braket{a | b}$. In the language of boundary CFT, $\ket{G_*}$ is an Ishibashi state \cite{Ishibashi1988, Cardy1989}. One should also note that by using this boundary condition we have implicitly assumed that $H_{\text{edge}}$ allows all quasiparticles, labelled by some vector in $\Gamma$, to tunnel across the real-space cut. Not allowing all such tunnelings would lead to $\ket{G_{\text{edge}}}$ having a structure calculated and detailed by Cano et al. \cite{Cano2015}, which leads to a ground state that cannot satisfy this boundary condition in the large real-space cut limit.  

Next, for a large but finite $L$ we must have that $S_b(L)$ will be the fixed point action plus \textit{irrelevant} perturbations. Thus, our correlators take the form $\bra{0} \phi_1(x_1, \tau_1) \phi_2(x_2, \tau_2)\dots e^{-\delta S_b(L)} \ket{G_*}$, where $\delta S_b(L)$ must be composed of irrelevant boundary perturbations. Eq. \ref{boundaryCon} indicates that all boundary operators correspond to local operators of the $A$ edge. Thus, $\delta S_b (L)$ must take the form,
\begin{equation}
    \delta S_b(L) = \sum_i \frac{\beta_i}{L^{h_i - 1}} \int dx' \phi_i (x')
\end{equation}
where $\phi_i(x')$ are local operators of the $A$ edge with corresponding scaling dimension $h_i$ and $\beta_i$ are numerical constants. The $L$ dependence of each term follows directly from the RG procedure that defines $S_b(L)$. As $\delta S_b(L)$ is an irrelevant perturbation, we must require that only operators with $h_i > 1$ appear. We use a prime on the $x$ coordinate as a reminder that this integral should be along the edge of the semi-infinite cylinder with circumference $L'$, as $S_b(L)$ is the boundary perturbation after the RG procedure. 

Finally, we thus have that so long as $\bra{a}\overline{\bra{b}}\ket{G_{\text{edge}}}$ corresponds to long wavelengths of a correlator we have,
\begin{equation} \label{overlapEq}
    \begin{split}
        \bra{a}\overline{\bra{b}}\ket{G_{\text{edge}}} &= \bra{a}\overline{\bra{b}}e^{-\delta S_b(L)} \ket{G_*} \\
        &= (\bra{a}e^{-\delta S_b(L)}) \overline{\bra{b}} \ket{G_*} \\
        &= \bra{a}e^{-\delta S_b(L)}\ket{b}
    \end{split} 
\end{equation}
where in going from the first to the second line we have used the fact $\delta S_b(L)$ is composed entirely of operators of the $A$ edge. Thus, we have found that overlaps with the ground state are encoded in a local boundary perturbation $\delta S_b(L)$.

\subsection{Entanglement spectrum} \label{ESasCFT}
Now we have determined the form of overlaps with the ground-state it is a rather simple matter to compute the reduced density matrix of $A$ and thus the ES. First, let $\{\ket{i}\}$ be an orthonormal basis for $A$ and let $\{ \overline{\ket{i}} \}$ be the corresonding basis of $B$. The reduced density matrix is simply given by,
\begin{equation}
    \begin{split}
        \rho_A &= \text{Tr}_B [ \ket{G_\text{edge}} \bra{G_\text{edge}}] \\
        &= \text{Tr}_B [ \sum_{ijkl} (\bra{i} e^{-\delta S_b (L)} \ket{j}) \ket{i} \overline{\ket{j}} \bra{k} \overline{\bra{l}} (\bra{l} e^{-\delta S_b^\dagger (L)} \ket{k}) ] \\
        &= e^{-\delta S_b (L)} e^{-\delta S_b^\dagger (L)}
    \end{split}
\end{equation}

We then define the entanglement action, $S_{ES}$ by $e^{-S_{ES}} = e^{-\delta S_b (L)} e^{-\delta S_b^\dagger (L)}$. As pointed out by DRR, we can use the Baker-Campbell-Hausdorff formula to express $S_{ES}$ as an expansion in commutators involving $\delta S_b(L)$ and $\delta S_b^{\color{blue} \dagger}(L)$. The key point is that the commutator of two operators that are an integrals of local operators, must itself be the integral of a local operator. We are thus led to the conclusion that the reduced density matrix of $A$ takes the form,
\begin{equation}
    \begin{split}
        \rho_A &= e^{-S_{ES}} \\
        S_{ES} &= \sum_i \frac{\alpha_i}{L^{h_i - 1}}\int dx' \phi_i(x')
    \end{split}
\end{equation}
where, once again, $\phi_i(x')$ are local operators of the $A$ edge with corresponding scaling dimension $h_i$ and $\alpha_i$ are numerical constants. The fact that $\delta S_b(L)$ has only operators with $h_i > 1$ implies that $S_{ES}$ must only contain terms with $h_i > 1$. Thus, we can interpret $S_{ES}$ as an \textit{irrelevant boundary perturbation} (in the RG sense). The fact that the reduced density matrix is Hermitian also implies that $S_{ES}$ is Hermitian ($S_{ES}^\dagger = S_{ES}$).

\section{Modeling the RSES of the Bosonic $\nu = 2/3$ Composite Fermion Wave Function} \label{ConstructionOf23Model}
We will now demonstrate how the general result of Sec. \ref{SurfaceCriticalTheory} can be applied to understand the RSES of the bosonic $\nu = 2/3$ Jain state. We choose to study this particular Jain state because it is potentially the simplest non-Laughlin state we can examine, both numerically and analytically. Specifically, we will develop a model for the RSES when this wave function is put on a sphere with $N$ particles and the real-space cut is along the equator. Firstly, the edge structure of this $\nu = 2/3$ state will be detailed, along with how certain observables, such angular momentum, are related to it. Then, a model is developed by arguing what edge operators should and should not appear in $S_{ES}$. The relation between our model and that proposed by Davenport et al. \cite{Davenport2015} will also briefly be discussed. We will then present a numerical test of this model in Sec. \ref{Section:NumericalTests}. 

\subsection{Edge structure of the $\nu = 2/3$ state} \label{EdgeTheory23}
It can be shown that the effective bulk Chern-Simons theory that one expects for a given CF state is equivalent to the effective bulk theory of the hierarchy state at the same filling fraction \cite{Read1990, Blok1990a}. Hence, the minimal edge theory of a given CF state should be the same as that for the corresponding hierarchy state.

The bosonic $\nu = 2/3$ hierarchy state can be thought of as a $\nu = 1/2$ droplet with a $\nu = 1/6$ droplet of quasi-particles. Following Wen \cite{Wen1995}, the edge theory can then be constructed as a $\nu = 1/2$ edge, described by a chiral boson field $\Tilde{\varphi}^{(1)}$, combined with a $\nu = 1/6$ edge described by another chiral boson field $\Tilde{\varphi}^{(2)}$. Following a Wick rotation and a mapping to the complex plane $z = e^{\frac{2\pi}{L}(\tau + ix)}$, the boson fields have the mode expansions,
\begin{equation}
    \Tilde{\varphi}^{(j)} = \Tilde{\varphi}^{(j)}_0 - i\Tilde{a}^{(j)}_0 \ln z + i\sum_{n \neq 0} \frac{1}{n} \Tilde{a}^{(j)}_n z^{-n}
\end{equation}
where,
\begin{equation}
    [\Tilde{a}^{(i)}_n, \Tilde{a}^{(j)}_m] = n \delta_{n+m, 0}\delta_{ij} \quad \quad [\Tilde{\varphi}^{(j)}, \Tilde{a}^{(k)}_0] = i \delta_{jk}
\end{equation}
and all other commutation relations are trivial. Further, the $a^{(i)}_0$ operators measure the $U(1)$ charge of their corresponding boson field.

The charge density operators, are $J^{(1)}(z) = i \frac{1}{2\pi\sqrt{2}}\partial \Tilde{\varphi}^{(1)}$ and $J^{(2)}(z) = i \frac{1}{2\pi\sqrt{6}}\partial \Tilde{\varphi}^{(2)}$ for the $\nu = 1/2$ and $\nu = 1/6$ edges respectively, where the charge is measured in units such that the underlying boson particles have charge one. Either edge can support excitations that have charge $1/2$. Given that the total charge of the combined edges must be an integer, the lattice, $\Gamma$, that defines this theory is generated by the basis vectors $\epsilon^1 = (\sqrt{2}, 0)$ and $\epsilon^2 = (1/\sqrt{2}, \sqrt{3/2})$.

The vacuum state, $\ket{0}$, is defined such that $\Tilde{a}^{(j)}_n \ket{0} = 0$ for all $n \geq 0$ and $j=1,2$. All states in the Hilbert space can then be generated by applying operators to the vacuum state, that are polynomials in $\Tilde{a}^{(j)}_{-n}$ ($n > 0$), $e^{i\sum_j \epsilon^1_j \Tilde{\varphi}^{(j)}_0}$ and  $e^{i\sum_j \epsilon^2_j \Tilde{\varphi}^{(j)}_0}$.

In order to numerically test the model we develop, we must be able to correctly label the angular momentum quantum number in our effective edge theory. In our case the edge is rotationally symmetric. In Appendix \ref{Appendix:AngularMomentum} we show that, under such circumstances, the angular momentum relative to the ground state (or relative to the edge vacuum $\ket{0}$), $\Delta M$, is given by,
\begin{equation} \label{AngularMomentumEq}
    \Delta M = L_0 + \sqrt{2}(Q^{(1)} - 1/2)\Tilde{a}_0^{(1)} + \sqrt{6} (Q^{(2)} - 1/2)\Tilde{a}_0^{(2)}
\end{equation}
where $Q^{(1)}$ and $Q^{(2)}$ are the amount of charge in the $\nu = 1/2$ and the $\nu = 1/6$ droplets respectively, and $L_0$ is the zeroth Virasoro mode which is given by $L_0 = \frac{1}{2}( (\Tilde{a}^{(1)}_0)^2 + (\Tilde{a}^{(2)}_0)^2 ) + \sum_{n > 0}( \Tilde{a}^{(1)}_{-n}\Tilde{a}^{(1)}_n + \Tilde{a}^{(2)}_{-n}\Tilde{a}^{(2)}_n)$. 

Finally, in the next subsection we will use a more convenient basis of fields given by,
\begin{equation}
    \begin{split}
        \varphi^{(1)} &= -\frac{\sqrt{3}}{2}\Tilde{\varphi}^{(1)} - \frac{1}{2} \Tilde{\varphi}^{(2)} \\
        \varphi^{(2)} &= \frac{1}{2}\Tilde{\varphi}^{(1)} - \frac{\sqrt{3}}{2} \Tilde{\varphi}^{(2)}
    \end{split}
\end{equation}
Let $\Delta Q^{(1)}$ and $\Delta Q^{(2)}$ be the electromagnetic charge added to the $\nu = 1/2$ and the $\nu = 1/6$ edges respectively. In this new basis of fields the $U(1)$ charge of $\varphi^{(1)}$ is given by $-\sqrt{\frac{3}{2}}( \Delta Q^{(1)} + \Delta Q^{(2)} )$ and the $U(1)$ charge of $\varphi^{(2)}$ is given by $\frac{1}{\sqrt{2}}( \Delta Q^{(1)} - 3\Delta Q^{(2)} )$. Thus, we can interpret the $\varphi^{(1)}$ as the charged mode and $\varphi^{(2)}$ as the neutral mode.

\subsection{The entanglement action} \label{Section:23ModelAction}
We can now apply the result of Sec. \ref{SurfaceCriticalTheory} to the specific case of $\nu = 2/3$ in the spherical geometry with the real-space cut along the equator, with $N$ total particles in the ground state. For the remainder of this paper we will use $N_A$ to denote the number of particles in subsystem $A$ (as is now a somewhat standard notation). In order to use our effective description we must specify what state of $A$ does our edge vacuum corresponds to, which is equivalent to specifying $Q^{(1)}$ and $Q^{(2)}$. For the results of Sec. \ref{SurfaceCriticalTheory} to apply we must pick the vacuum state such that it corresponds to the state with $\textit{minimum}$ entanglement pseudo-energy in hemisphere $A$. Such a state will have $N/2$ particles in $A$. If we let $\Delta N_A$ be the change in the number of particles in $A$ from this minimum state, then this choice of vacuum means that the $U(1)$ charge of $\varphi^{(1)}$ is given by $-\sqrt{\frac{3}{2}}\Delta N_A$. In general, this vacuum state does not correspond to the lowest angular momentum eigenstate of the reduced density matrix of $A$ in the $N_A = N/2$ sector. If we wish to compare this model to the numerically calculated RSES we need to understand how this effective theory behaves near this lowest angular momentum state. In what follows we shall first discuss what form $S_{ES}$ takes for the minimum pseudo-energy vacuum choice and then we shall detail how this can be transformed to give an effective description around the lowest angular momentum state.

The local operators that can appear in $S_{ES}$ can be labelled by two tuples of positive integers, $\mathbf{k}^{(1)}$ and $\mathbf{k}^{(2)}$ with $\mathbf{k}^{(i)} = (k^{(i)}_1, k^{(i)}_2, \dots, k^{(i)}_{p_i})$, where the corresponding operator is given by,
\begin{equation}
    \phi_{(\mathbf{k}^{(1)}, \mathbf{k}^{(2)})} (z) = :\prod_{r = 1}^{p_1}i\partial^{k^{(1)}_r} \varphi^{(1)}(z) \prod_{s = 1}^{p_2} i\partial^{k^{(2)}_s} \varphi^{(2)}(z):
\end{equation}
Where $:*:$ indicates normal ordering. One might also worry about vertex terms, $e^{ia\varphi^{(1)} + ib\varphi^{(2)}}$. Vertex terms in $\varphi^{(1)}$ are strictly forbidden, as $S_{ES}$ must conserve the electromagnetic charge. Further, there is reason to believe that vertex operators involving $\varphi
^{(2)}$ should not appear in $S_{ES}$. Firstly, by directly modeling the interactions between the two edges, Cano et al. predicted an $S_{ES}$ that does not include vertex terms (at leading order) \cite{Cano2015}. Secondly, the Wen-Zee effective Chern-Simons theory indicates there are two emergent conserved $U(1)$ currents at very low energy and long length scales\cite{Wen1992}. The $U(1)$ charges of our two edge modes are supposed to correspond with the charges of these currents. Thus, if these two $U(1)$ charges are seperately conserved, then $S_{ES}$ should conserve both $U(1)$ charges of our edge modes. Hence, we do not expect vertex terms to appear and we will assume this to be the case at least to the orders in $1/\sqrt{N}$ that we are considering \footnote{It was also tested numerically if adding vertex terms in $S_{ES}$ would improve the fit to numerical data. However, it was found that such terms made no difference to the quality of the fit and the fitted coefficients of such terms had confidence intervals that overlaped with zero.}.

The scaling dimension of each operator is simply given by $h_{(\mathbf{k}^{(1)}, \mathbf{k}^{(2)})} = \sum_{r=1}^{p_1} k^{(1)}_r + \sum_{s=1}^{p_2} k^{(2)}_s$. Once again, the condition that $S_{ES}$ is an irrelevant boundary perturbation means that only operators with $h_{(\mathbf{k}^{(1)}, \mathbf{k}^{(2)})} > 1$ can appear.

Noting that with the real-space cut on the equator $L \propto \sqrt{N}$, we can express $S_{ES}$ in radial quantisation coordinates, $z = e^{\frac{2 \pi }{L} (\tau + ix)}$, as,
\begin{equation} \label{masterS}
    S_{ES} = \sum_{j} \frac{K_j}{\sqrt{N}^{h_j - 1}}\oint \frac{dz}{2\pi i} z^{h_j - 1} \phi_j (z)
\end{equation}
where $j$ is a shorthand notation for $(\mathbf{k}^{(1)}, \mathbf{k}^{(2)})$ and $K_j$ are numerical constants. Note that the operators that result from integrating these local operators, $\phi_j(z)$, are not linearly independent \cite{Fern2018a, Fern2018}. For example, through integration by parts we have $\oint dz z :\partial^2 \varphi^{(2)}(z): = - \oint dz :\partial \varphi^{(2)}(z):$. Hence, if we pick a particular basis of operators the coefficient of each term will generically not just scale as $1/\sqrt{N}^{h_j - 1}$, but will be a sum of different powers: $1/\sqrt{N}^{h_j - 1} + a/\sqrt{N}^{h_j} + b/\sqrt{N}^{h_j + 1} + \dots$.

One can see from Eq. \ref{masterS} that for very large system sizes we can approximate $S_{ES}$ with an action that only contains terms with scaling dimensions below some small integer, and hence a finite number of terms. Our intention is to test this local entanglement action to as high an order as possible.

Generally, at scaling dimension $n$ the number of possible operators (when not considering linear independence after integration) is $\sum_{i=0}^n$ ( \# partitions of $i$ ) $\times$ (\# partitions of $n - i$), which grows very quickly with $n$. To test any truncated entanglement action can then involve many free parameters, which can then become unpractical very quickly at higher order. This is why we have decided to test the entanglement action in the case of the real-space cut being the equator of a sphere, as this introduces a \textit{symmetry} that can cut down the number of possible operators. In Appendix \ref{Appendix:Symmetry} we briefly consider the case where the cut is not along 
the equator. On the sphere with a cut along the equator, the system is symmetric under exchange of the two hemispheres, $A$ and $B$. This symmetry between $A$ and $B$ implies that the RSES must be invariant under $\Delta N_A \rightarrow - \Delta N_A$. It is not obvious how this symmetry is manifest in our effective description. In Appendix \ref{Appendix:Symmetry} we argue, although not entirely rigorously, that this transformation in the effective description is given by, $i\partial\varphi^{(1)} \rightarrow -i\partial\varphi^{(1)}$ and $i\partial\varphi^{(2)} \rightarrow i\partial\varphi^{(2)}$. At the very least the quality of the fit of our resulting model to the numerical RSESs in Sec. \ref{Section:NumericalTests} give some support to this not-entirely-rigorous argument. Furthermore, we give another numerical test of this symmetry in Appendix \ref{Appendix:Symmetry}. Assuming this symmetry gives a further restriction on the allowed operators in that $\mathbf{k}^{(1)}$ must have an even number of components (i.e. $p_1$ must be even). This then cuts down the number of possible of operators, which will allow us to test for the local entanglement action in a practical manner.

We shall now present two approximations of $S_{ES}$: one with terms only at scaling dimension 2, $S^{\{ 2 \}}_{ES}$, and one with terms at scaling dimension 3 and lower $S^{\{ 3 \} }_{ES}$.

At scaling dimension 2 the allowed operators are, $:(i\partial \varphi^{(1)})^2:$, $:(i\partial \varphi^{(2)})^2:$ and $:i\partial^2 \varphi^{(2)}:$. Hence, $S^{\{ 2 \}}_{ES}$ will take the form,
\begin{equation}
    \begin{split}
        S^{\{ 2 \}}_{ES} =& \oint \frac{dz}{2 \pi i} z \bigg [ \alpha :i\partial^2 \varphi^{(2)}: \\
        &+ \frac{\beta}{2} :(i\partial \varphi^{(1)})^2: + \frac{\gamma}{2} :(i\partial \varphi^{(2)})^2: \bigg ]
    \end{split} 
\end{equation}
where $\alpha$, $\beta$ and $\gamma$ are numerical coefficients that depend on $N$. All three terms can be generated through integration by parts from terms of higher scaling dimension and hence the $N$ dependence of each coefficient should take the form $\frac{a}{\sqrt{N}} + \frac{b}{N} + \dots$.

At scaling dimension 3 the allowed operators, that are not related to any at scaling dimension 2 by integration by parts, are $:(i\partial \varphi^{(2)})^3:$ and $:(i\partial \varphi^{(2)})(i\partial \varphi^{(1)})^2:$. We then have $S^{\{ 3 \}}_{ES}$ taking the form,
\begin{equation}
    \begin{split}
        S^{\{ 3 \}}_{ES} =& S^{\{ 2 \}}_{ES} + \oint \frac{dz}{2 \pi i} z^2 \bigg [ \delta :(i\partial \varphi^{(2)})^3: \\
        &+ \epsilon :(i\partial \varphi^{(2)})(i\partial \varphi^{(1)})^2: \bigg ]
    \end{split}
\end{equation}
Both these terms can also be generated through integration by parts of higher scaling dimension operators. Thus, the $N$ dependence of $\delta$ and $\epsilon$ will take the form $\frac{a}{N} + \frac{b}{N\sqrt{N}} + \dots$.

As mentioned at the start of this section, when we test these approximate entanglement actions we are limited to comparing to the numerically calculated spectrum at angular momenta near the lowest possible for a particular $\Delta N_A$ sector. In Appendix \ref{Appendix:AngularMomentum} we show that, in the $\Delta N_A = 0$ sector, the lowest angular momentum state and the state with lowest entanglement pseudo-energy differ in angular momentum by $O(N)$ (in big O notation). In our effective description this lowest angular momentum state, $\ket{\hat{0}}$, must correspond to a state which is just some $U(1)$ charge of the neutral mode added to the vacuum, $\ket{\hat{0}} = e^{i\kappa \varphi^{(2)}_0} \ket{0}$ for some $\kappa$. As the angular momentum depends on the charge of the neutral mode quadratically, we have that $\kappa = O(\sqrt{N})$.

We then define a new neutral mode with a shifted $U(1)$ charge, $i\partial \hat{\varphi}^{(2)}(z) \equiv i \partial \varphi^{(2)}(z) - \frac{\kappa}{z}$. The lowest angular momentum state, $\ket{\hat{0}}$, will be the vacuum of these shifted fields. We can rewrite $S_{ES}$ in terms of this new neutral mode, which will merely ``reshuffle'' the various terms. For example, $z:(i\partial \varphi^{(2)})^2: = z:(i\partial \hat{\varphi}^{(2)})^2: - 2\kappa i\partial \hat{\varphi}^{(2)} + \frac{\kappa^2}{z}$. In terms of the shifted neutral mode the approximate entanglement actions will take the form,
\begin{equation} \label{d2S}
    \begin{split}
        \hat{S}^{\{ 2 \}}_{ES} =& \oint \frac{dz}{2 \pi i} \hat{\alpha} :i\partial \hat{\varphi}^{(2)}: \\
        &+ z \bigg [ \frac{\hat{\beta}}{2} :(i\partial \varphi^{(1)})^2: + \frac{\hat{\gamma}}{2} :(i\partial \hat{\varphi}^{(2)})^2: \bigg ]
    \end{split} 
\end{equation}
and
\begin{equation} \label{d3S}
    \begin{split}
        \hat{S}^{\{ 3 \}}_{ES} =& \hat{S}^{\{ 2 \}}_{ES} + \oint \frac{dz}{2 \pi i} z^2 \bigg [ \hat{\delta} :(i\partial \hat{\varphi}^{(2)})^3: \\
        &+ \hat{\epsilon} :(i\partial \hat{\varphi}^{(2)})(i\partial \varphi^{(1)})^2: \bigg ]
    \end{split}
\end{equation}
where we have now integrated the $z:i\partial^2 \hat{\varphi}^{(2)}:$ term by parts to give $:i\partial \hat{\varphi}^{(2)}:$. Note the coefficients $\hat{\alpha}, \hat{\beta}, \hat{\gamma}, \hat{\delta}$ and $\hat{\epsilon}$ will be different from $\alpha, \beta, \gamma, \delta$ and $\epsilon$, not just by reshuffling terms already present in $S^{\{ 2 \}}_{ES}$ and $S^{\{ 3 \}}_{ES}$, but also by the reshuffling of higher order terms in the full entanglement action, $S_{ES}$. The $N$ dependence of $\hat{\beta}, \hat{\gamma}, \hat{\delta}$ and $\hat{\epsilon}$ will take the same form as before. However, the $N$ dependence of $\hat{\alpha}$ now takes the form: constant$ + \frac{a}{\sqrt{N}} + \frac{b}{N} + \dots$. The constant comes from the fact that  $z:(i\partial \varphi^{(2)})^2: = z:(i\partial \hat{\varphi}^{(2)})^2: - 2\kappa i\partial \hat{\varphi}^{(2)} + \frac{\kappa^2}{z}$ and $\kappa = O(\sqrt{N})$.

Eqs. \ref{d2S} and \ref{d3S} define the two models we shall test against the numerically calculated RSES, where the parameters of each model cannot easily be calculated from the microscopic wave function and instead must be determined through a fitting procedure. In Appendix \ref{Appendix:Entropy} we show that this model, based on an entanglement action, reproduces the expected entanglement entropy for this quantum Hall state, which some readers may find useful.

In the case of the $S^{\{2 \}}_{ES}$ model, for a \textit{fixed} $N_A$ the entanglement levels can be labelled by an integer $n \in \mathbb{Z}$ and two partitions $\lambda^{(1)}$ and $\lambda^{(2)}$, with numerical value $\Delta \xi_{n, \lambda^{(1)}, \lambda^{(2)}} = \sqrt{2}\alpha n + \beta |\lambda^{(1)}| + \gamma |\lambda^{(2)}| +$ constant, where $|\lambda^{(i)}| = \sum_{m \in \lambda^{(i)}} m$. For this fixed $N_A$ one can see that this spectrum is equiveleant to the spectrum of the fermion Hamiltonian, $H = \sum_m[(\beta m + \alpha/\sqrt{2})c^\dagger_{m,1}c_{m,1} + (\gamma m - \alpha/\sqrt{2})c^\dagger_{m,2}c_{m,2}]$ with a fixed number of fermions, where $c^\dagger_{m,i}$ and $c_{m,i}$ are fermionic creation and annihilation operators respectively. Hence, for a fixed $N_A$ the model presented here and the model proposed by Davenport et al. \cite{Davenport2015} are equivalent. However, the two models cannot be equivalent over multiple $N_A$ sectors of the RSES (as mentioned in the introduction).

\section{Numerical Tests} \label{Section:NumericalTests}
We now present two numerical tests for our effective model of the RSES of the bosonic $\nu = 2/3$ composite fermion wave function on the sphere with the real-space cut on the equator. The first test demonstrates how well the effective description can be fit to the numerically computed RSES, over multiple $N_A$ sectors. We then test how the fitted model parameters vary with the system size.

\subsection{Fitting the model} \label{modelfittest}
In all cases presented here, the RSES has been calculated numerically using the methods of Rodriguez et al. \cite{Rodriguez2013}, where the projection to the lowest Landau level has been performed using the Jain-Kamila procedure \cite{Jain1997} and inner products have been calculated using Monte-Carlo integration (see Appendix \ref{Appendix:Numerics} for details). The parameters in the model entanglement actions, $S^{\{2\}}_{ES}$ and  $S^{\{3\}}_{ES}$, are then fit using a least squares procedure, which is detailed in Appendix \ref{Appendix:FittingProcedureAndFittedValues} along with the procedure used for error estimation. One subtlety that should be mentioned, is that the least squares fitting procedure used here cannot determine the sign of the parameters $\hat{\alpha}$, $\hat{\delta}$ and $\hat{\epsilon}$, as simultaneously changing the sign of these parameters will produce the same spectrum. The sign of such parameters would have to be determined by some other means.

\begin{figure*}
    \centering
    \includegraphics{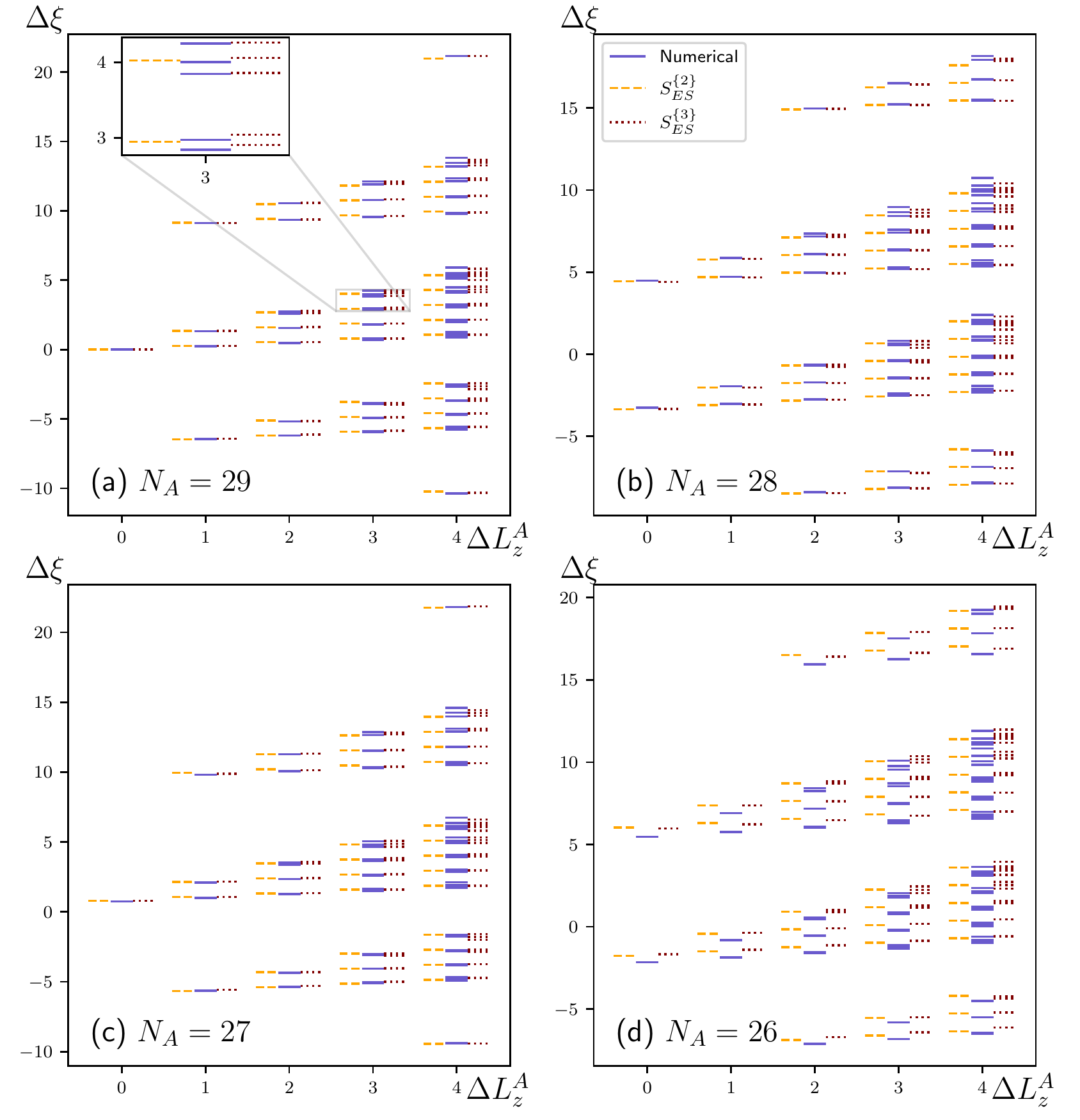}
    \caption{Comparison between the numerically calculated RSES (blue-solid), using the methods of \cite{Rodriguez2013} (see Appendix \ref{Appendix:Numerics} for details), and the models, $S^{\{ 2 \}}_{ES}$ (Eq. \ref{d2S}) (yellow-dashed) and $S^{\{ 3 \}}_{ES}$ (Eq. \ref{d3S}) (red-dotted), fitted using the procedure of Appendix \ref{Appendix:FittingProcedureAndFittedValues}, in the case of $N = 58$ bosons on a sphere, in the $\nu = 2/3$ bosonic Jain wave function with real space cut on the equator over the (a) $N_A = 29$, (b) $N_A = 28$, (c) $N_A = 27$ and (d) $N_A = 26$ sectors. $\Delta L_z^A$ is the angular momentum relative to the lowest possible for the given $N_A$ sector and $\Delta \xi$ is the entanglement pseudo-energy relative the level at $N_A = 29$ and $\Delta L_z^A = 0$. Error bars have not been included, as they are very small and would not be easily visible on this plot.}
    \label{fig:fittingTest}
\end{figure*}

The resulting numerically calculated and fitted RSES for the case of $N = 58$ bosons can be seen in Fig. \ref{fig:fittingTest}, where the spectra are shown over the $N_A= 29, 28, 27, 26$ sectors.
One can see that the fitted low order entanglement action, $S^{\{2\}}_{ES}$, is in good agreement with the numerically calculated spectrum over the $N_A = 29, 28, 27$ sectors. There is an improvement by adding the higher order terms of $S^{\{3\}}_{ES}$, where we can see some of the higher order level splitting being reproduced. One can also see that at higher angular momentum both effective models, $S^{\{ 2 \}}_{ES}$ and $S^{\{ 3 \}}_{ES}$, begin to break down. This is due to the presence of the higher order terms in the entanglement action whose coefficients are small, but have matrix elements which grow much faster, as momentum is increased, compared to the lower order terms included in both models. This also explains the observed deviation at $N_A = 26$, as at scaling dimension 4 one would expect the $:(\partial \varphi^{(1)})^4:$ term to be present which would give a $(a_0^{(1)})^4 \propto (\Delta N)^4$. This term is not present in either model and would lead to their suddenly large inaccuracy at larger $\Delta N_A$.

\subsection{Parameter scaling} \label{Section:ScaleTest}
\begin{figure}[h!]
    \centering
    \includegraphics{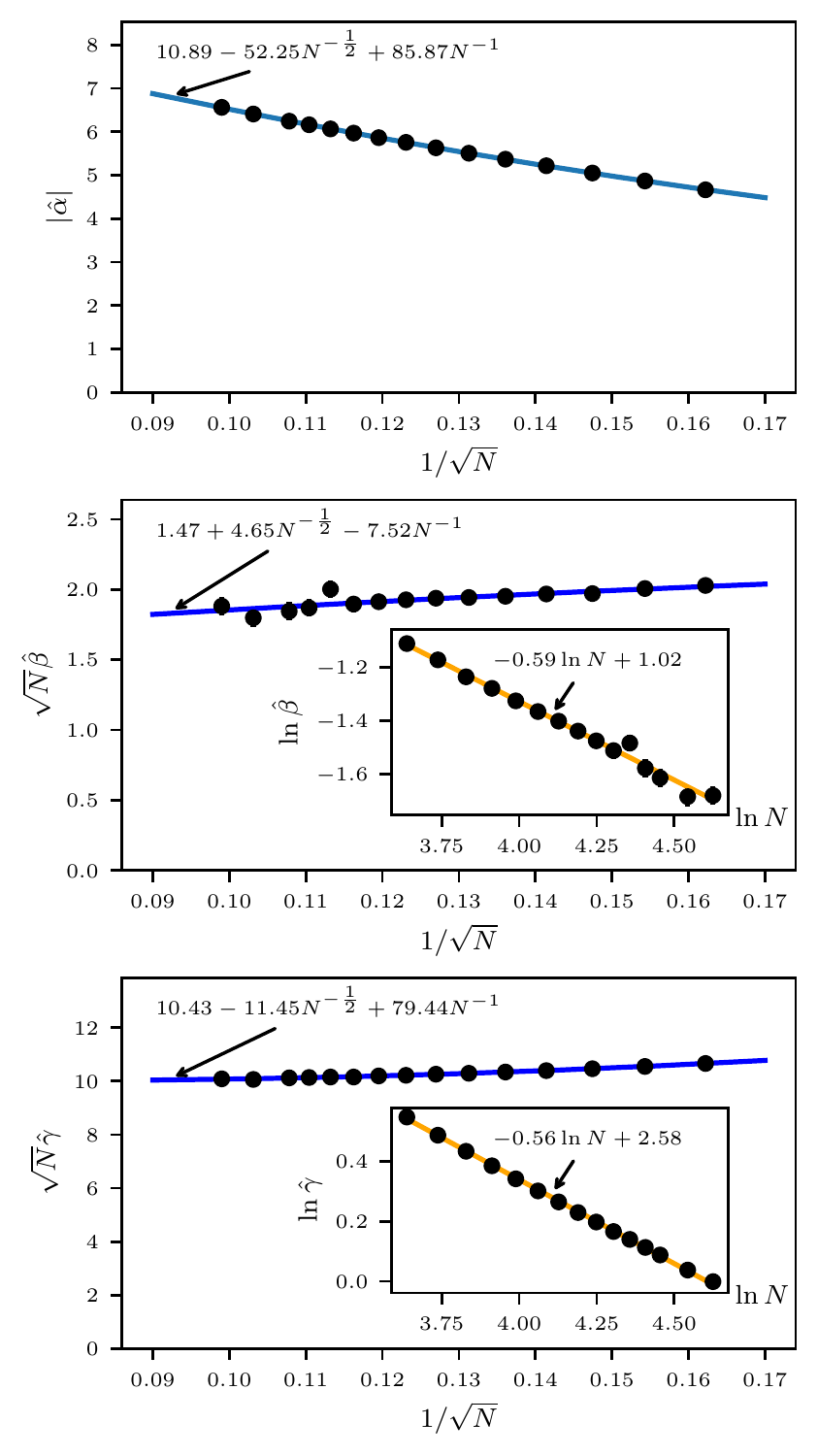}
    \caption{The $N$ dependence of the fitted $\hat{\alpha}$, $\hat{\beta}$ and $\hat{\gamma}$ parameters, in the $S^{\{ 2 \}}_{ES}$ model (Eq. \ref{d2S}) of the $\nu = 2/3$ bosonic Jain wave function RSES. There is a sign ambiguity when fitting the parameter $\hat{\alpha}$, and hence we have only presented its absolute value (see Sec. \ref{Section:ScaleTest}). The $N$ dependence of all three parameters can be fit to the expected functional form (blue line) (see discussion of Sec. \ref{ConstructionOf23Model}). The insets show the $\ln N$ dependence of $\ln \hat{\beta}$ and $\ln \hat{\gamma}$ with corresponding linear regressions (orange line), which indicate a leading order scaling close to $N^{-1/2}$ for both $\hat{\beta}$ and $\hat{\gamma}$. Error bars have been included, but in many cases are not visible due to some errors being very small. Our procedures for error estimation is given in Appendix \ref{Appendix:FittingProcedureAndFittedValues}. }
    \label{fig:scale2Plot}
\end{figure}

\begin{figure}[h!]
    \centering
    \includegraphics{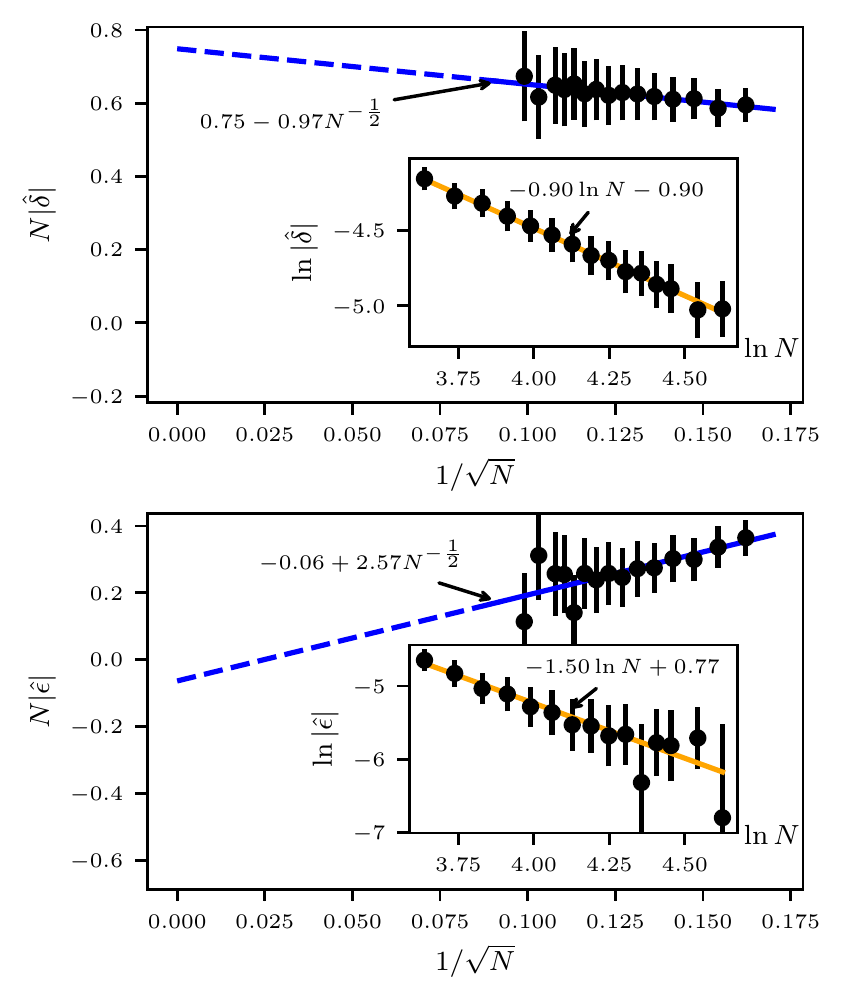}
    \caption{Same as for Fig. \ref{fig:scale2Plot}, but now for the $\hat{\delta}$ and $\hat{\epsilon}$ parameters in the $S^{\{ 3 \}}_{ES}$ model. There is a sign ambiguity when fitting the parameters $\hat{\delta}$ and $\hat{\epsilon}$, and hence we have only presented their absolute values (see Sec. \ref{Section:ScaleTest}). $\hat{\delta}$ can be fit to its expected $N$ dependence with the $\ln \hat{\delta}$ inset indicating a leading order scaling close to the expected $N^{-1}$. $\hat{\epsilon}$ can also be fit to it's exptected $N$ dependence and the $\ln \hat{\epsilon}$ inset indicates a leading order scaling of $N^{-3/2}$. This is still consistent with the predictions of Sec. \ref{ConstructionOf23Model} (see section \ref{Section:ScaleTest}). Our procedures for error estimation is given in Appendix \ref{Appendix:FittingProcedureAndFittedValues}.}
    \label{fig:scale3Plot}
\end{figure}
To test how the parameters of each model vary with the system size, $N$, we have repeated the procedure of Sec. \ref{modelfittest} to estimate these parameters for a  variety of system sizes between $N = 38$ and $N = 102$. One should keep in mind again that the sign of parameters $\hat{\alpha}$, $\hat{\delta}$ and $\hat{\epsilon}$ cannot be determined by the fitting procedure used; we have thus taken the absolute value of any parameter values presented in this section. As discussed in Sec. \ref{ConstructionOf23Model} each parameter should vary with $N$ as, $a/\sqrt{N}^{h-1} + b/\sqrt{N}^{h} + c/\sqrt{N}^{h + 1} + \dots$, where $h$ is the scaling dimension of the term in the entanglement action which this particular parameter is a coefficient of. We have then fitted the $N$ dependence of each parameter to this functional form (whilst keeping $h$ fixed to the expected value). For each parameter which decays to zero for large $N$, we have also performed a linear regression for the $\ln N$ dependence of the logarithm of the parameter. This allows us to check that the fit to the expected $N$ dependence has not been too biased, by estimating the exponent of the leading order contribution.

The result of this test for the parameters of $S^{\{2 \}}_{ES}$ can be seen in Fig. \ref{fig:scale2Plot}. One can see that all three parameters can be fit very well to their expected $N$ dependence. Furthermore, the linear regressions of $\ln \hat{\beta}$ and $\ln \hat{\gamma}$ vs. $\ln N$ both indicate $\hat{\beta} \sim N^{-0.59}$ and $\hat{\gamma} \sim N^{-0.56}$, which is close to the expected leading order scaling of $N^{-1/2}$. One should note that the presence of the higher order terms in $1/\sqrt{N}$ will cause the estimation of this leading order exponent to be slightly off and one can also visually see the presence of such terms from the curvature in the $\ln \hat{\gamma}$ plot. 

Fig. \ref{fig:scale3Plot} shows the result of this test for the parameters of $S^{\{ 3 \}}_{ES}$, where we have not shown the results for $\hat{\alpha}$, $\hat{\beta}$ and $\hat{\gamma}$ here as these are nearly identical to those in the $S^{\{ 2 \}}_{ES}$ case (as one would expect) and can be seen in Appendix \ref{Appendix:FurtherPlots}. The dependence of $\hat{\delta}$ of $N$ can be fit well to the functional form and the estimation of the leading order exponent gives $\hat{\delta} \sim N^{-0.90}$, which is close to the expected leading dependence, $N^{-1}$. It can be seen that $\hat{\epsilon}$ can be fit to it's expected $N$ dependence. However, the estimation of the leading order exponent indicates that the fitted $\hat{\epsilon}$ will scale as $\sim N^{-3/2}$ at large system size, which is also evident in the small $N\hat{\epsilon}$-axis intercept of the fitted $N$ dependence. The scaling of $\hat{\epsilon}$ is still consistent with the predictions of Sec. \ref{ConstructionOf23Model}. We have that, through integration by parts, $-\oint dz z^2 :(i\partial \hat{\varphi}^{(2)})(i\partial \varphi^{(1)})^2: = \oint \frac{z^3}{3} [ :(i\partial^2 \hat{\varphi}^{(2)})(i\partial \varphi^{(1)})^2: + :(i\partial \hat{\varphi}^{(2)}) \partial (i\partial \varphi^{(1)})^2:$. Hence, the $\hat{\epsilon}$ term can be generated by scaling dimension 4 terms. In general, $\hat{\epsilon}$ can then be expanded as $\hat{\epsilon} = a/N  + b/N^{3/2} + \dots$. If it is the case that $a = 0$ then we would have $\hat{\epsilon}$ would scale as $\hat{\epsilon} \sim N^{-3/2}$.

\section{Conclusion} \label{Conclusion}
We have studied the real-space entanglement spectrum (RSES) of composite fermion wave functions. Starting from the Qi-Katsura-Ludwig \cite{Qi2012} cut and glue approach, we argued and numerically substantiated that the real-space entanglement spectrum of fully chiral abelian quantum Hall states is given by the spectrum of a local boundary perturbation of a $(1+1)$d conformal field theory (CFT), which describes an effective edge dynamics along the real-space cut. The CFT was assumed to be that which describes the minimal edge of the corresponding quantum Hall state (i.e. the edge in the absence of reconstruction). This is the Dubail-Read-Rezayi (DRR) ``scaling property'' of the RSES \cite{Dubail2012}.

The cut and glue approach gives the starting assumption that the low-lying RSES is equivalent to the entanglement spectrum between a chiral and anti-chiral quantum Hall edge in the ground state, $\ket{G_{\text{edge}}}$, of some ``physical'' gapped Hamiltonian only involving the edge degrees of freedom, $H_{\text{edge}}$. We then converted the problem of computing overlaps with $\ket{G_{\text{edge}}}$ to a boundary critical problem, where the general form of the overlaps could then be determined by standard renormalization group arguments combined with the methods of boundary conformal field theory. From the understanding of these overlaps the scaling property of the RSES then simply followed. This scaling property implies that at large real-space cut length, one only needs to use terms of a low scaling dimension from the CFT to accurately reproduce the RSES. 

We then used this result to develop a model for the RSES of the bosonic $\nu = 2/3$ Jain wave function on the sphere with the real-space cut along the equator. Two numerical tests were presented for this model: one demonstrating the model can be fit to the numerically calculated RSES well, and one which demonstrated the fitted coefficients of the various local operators scaled with the system size as predicted by our renormalization group arguments. 

The success of our numerical tests hints that it may be possible to extend the original arguments of DRR to the CF wave functions, which as a byproduct would also give an extension of Laughlin's plasma analogy \cite{Laughlin1983} to such wave functions. In their work, DRR showed for a large class of trial wave functions, the scaling property of the RSES holds under the assumption of the ``generalized screening'' hypothesis. This screening hypothesis is an extension of the Laughlin plasma analogy to many other trial wave functions \cite{Read2009}. These trial wave functions are those that can be directly written as correlation functions of some CFT (conformal blocks). In contrast, CF wave functions can be expressed as appropriately symmetrized conformal blocks, but cannot be directly written as conformal blocks, as mentioned in the introduction \cite{Hansson}. Thus, to extend DRR's work one must work out if this screening hypothesis has an analogous form when dealing with symmetrized conformal blocks. 

This screening hypothesis is what allowed DRR to proceed in such a rigorous manner and allowed them to understand much more about these states as a by-product, such as ``edge-state'' inner products and a precise bulk-edge correspondence. The conformal block structure is also intimately related to the fact these trial wave functions have an associated ``special'' parent Hamiltonian, for which the corresponding trial wave function and quasi-hole states are exact zero energy states \cite{Jackson2013a}. The existence of this ``special'' parent Hamiltonian allows one to understand a large amount of the physics of the corresponding phase of matter \cite{Simon2020}. While such Hamiltonians have been shown to exist for the \textit{unprojected} CF wave functions \cite{Bandyopadhyay2020, Greiter2021}, 
it is still not known any exist for the projected CF wave functions \cite{Sreejith2018}. Thus, it is still not known, in general, if much of the same physics still applies to the CF wave functions, which is well understood for trial wave functions that do have ``special'' parent Hamiltonians, as mentioned in the introduction. Here, at least, we have demonstrated the DRR scaling property of the RSES applies.

\begin{acknowledgments}
SHS is supported by EPSRC Grant EP/S020527/1. Statement of
compliance with EPSRC policy framework on research
data: This publication is theoretical work that does not
require supporting research data.
Part of the calculations in this work was performed using computing resources procured using DST/SERB grant ECR/2018/001781.
We acknowledge National Supercomputing Mission (NSM) for providing computing resources of PARAM Brahma at IISER Pune, which is implemented by C-DAC and supported by the Ministry of Electronics and Information Technology (MeitY) and Department of Science and Technology (DST), Government of India. 
\end{acknowledgments}

\appendix

\section{Hemisphere Swapping Symmetry} \label{Appendix:Symmetry}
In Sec. \ref{ConstructionOf23Model} we mentioned that $S_{ES}$, for the $\nu = 2/3$ state with a cut along the equator, has a symmetry due to the fact the RSES should be invariant under swapping the two hemispheres. We claimed that this symmetry implies that $S_{ES}$ should be invariant under $i\partial \varphi^{(1)} \rightarrow -i\partial \varphi^{(1)}$, $i\partial \varphi^{(2)} \rightarrow i\partial \varphi^{(2)}$. Whilst we do not have a rigorous proof for this statement, we wish to give a strong motivation for it. We shall first detail how this symmetry is manifest in the $\nu = 2$ RSES. Then, with the knowledge of the $\nu = 2$ RSES, we deduce how this symmetry should be manifest in the $\nu = 2/3$ state using the composite fermion description. Finaly, we will present a simple numerical test for this symmetry.

\subsection{The $\nu = 2$ RSES on the sphere}
On the sphere with $N_\phi = 2Q$ flux quanta passing through it, the first Landau level orbitals, $\psi_{1,m}$, and the second Landau level orbitals, $\psi_{2,m}$, take the form,
\begin{equation}
    \begin{split}
        \psi_{1,m} (u,v) =& \mathcal{N}_{1, m} v^{Q - m} u^{Q + m} \\
        m =& -Q, -Q + 1, \dots, Q - 1, Q \\
        \\
        \psi_{2, m} (u,v) =& \mathcal{N}_{2, m} v^{Q - m} u^{Q + m}  \\
        &\times [ (Q + 1 + m)v^* v - (Q + 1 - m)u^* u] \\
        m =& -Q - 1, -Q, \dots, Q, Q + 1 \\
    \end{split}
\end{equation}
where $u = \cos (\theta /2)e^{i\phi/2}$, $v = \sin (\theta /2 ) e^{-\phi /2}$ and $\mathcal{N}_{i,m}$ are normalisation factors \cite{Haldane1983, Jain2007}. The wave function of the full $\nu = 2$ state can be written in second quantised notation as,
\begin{equation}
    \ket{\nu = 2} = \prod_{m = -Q}^Q c^\dagger_{1,m} \prod_{n = -Q-1}^{Q+1} c^\dagger_{2, n} \ket{0}
\end{equation}

To compute the RSES one must rewrite this state in terms of orbitals that are orthogonal on the $A$ and $B$ hemispheres separately \cite{Peschel2003, Davenport2015}. We then define the correlation matrix to be,
\begin{equation}
    C^{A,m}_{ij} = \int_A d\Omega \psi^*_{i,m} (u,v) \psi_{j, m} (u,v)
\end{equation}
Further, we then define $\psi_{+,m}$ and $\psi_{-,m}$ to be the orbitals that correspond to the highest eigenvalue, $\lambda^{A,m}_+$ and the lowest eigenvalue $\lambda^{A,m}_-$ of the correlation matrix respectively. As these orbitals diagonalise the correlation matrix they must be orthogonal on the $A$ and $B$ hemispheres separately.

One can then verify that the $\nu = 2$ state can be written in terms of these orbitals simply as,
\begin{equation}
    \ket{\nu = 2} = \prod_m c^\dagger_{+,m} \prod_n c^\dagger_{-,m} \ket{0}
\end{equation}

By writing $c_{\pm, m} = \sqrt{\lambda^{A,m}_\pm} c^A_{\pm, m} + \sqrt{1-\lambda^{A,m}_\pm} c^B_{\pm, m}$, where $c^A_{\pm, m}$ and $c^B_{\pm, m}$ is the $c_{\pm, m}$ orbital restricted to the $A$ and $B$ hemispheres respectively, one can then Schmidt decompose the state $\ket{\nu = 2}$. The resulting entanglement Hamiltonian for $A$ is,
\begin{equation}
    S^A_{ES} = \sum_{\pm, m} \ln \bigg [ \frac{1-\lambda^{A,m}_\pm}{\lambda^{A,m}_\pm} \bigg ] c^{A\dagger}_{\pm, m} c^{A}_{\pm, m}
\end{equation}

Under the transformation $u \rightarrow v$, $v \rightarrow u$, the Landau level orbitals transform as $\psi_{1,m} \rightarrow \psi_{1,-m}$ and $\psi_{2,m} \rightarrow -\psi_{2,-m}$. By first considering the transformation properties of the correlation matrix, one can also show under this transformation $\psi_{\pm, m} \rightarrow -\psi_{\mp, -m}$. Furthermore, the transformation properties of the correlation matrix gives, $\lambda^{A,m}_\pm = \lambda^{B,-m}_\pm$. Note that an eigen-orbital of the $A$ correlation matrix with the lowest eigenvalue, at a particular angular momentum, must be an eigen-orbital of the $B$ correlation matrix with the highest eigenvalue, at the same angular momentum (i.e. the orbital $\psi_{\pm, m}$ will have eigenvalue $\lambda^{B,m}_\mp$ for the corresponding $B$ correlation matrix). Finally, we note that under this transformation an orbital entirely on the $A$ hemisphere will be mapped to an orbital entirely on the $B$ hemisphere, so we have $c^{A\dagger}_{\pm, m} c^{A}_{\pm, m} \rightarrow c^{B\dagger}_{\mp, -m} c^{B}_{\mp, -m}$. Putting this altogether, one finds that this transformation gives $S^A_{ES} \rightarrow S^B_{ES}$. This should be the case as under this transformation our entire wave function is invariant.

Due to the fact that the number of electrons in each orbital are separately conserved (for this non-interacting $\nu = 2$ state) we also have that under $c^{B\dagger}_{\pm, m} c^{B}_{\pm, m} \rightarrow 1 - c^{A\dagger}_{\pm, m} c^{A}_{\pm, m}$ we have $S^B_{ES} \rightarrow S^A_{ES}$ up to an additive constant. Putting these two transformations together maps $S^A_{ES}$ onto itself. 
We can deduce these properties from a more cartoon picture in the effective description, which will be more helpful in dealing with the $\nu = 2/3$ case. Let $\partial_x \phi_\pm$ represent the electron density in the $\pm$ orbitals at the edge. As under the hemisphere swapping transformation $\psi_{\pm, m} \rightarrow \psi_{\mp, -m}$, we must have $\partial_x \phi_\pm \rightarrow \partial_x \phi_\mp$ under this transformation. Thus, as our entire wave function is invariant under this transformation we must have $S^A_{ES}(\partial_x \phi_+ , \partial_x \phi_-) = S^B_{ES}(\partial_x \phi_-, \partial_x \phi_+)$. Further, as the number of electrons in either orbital is locally conserved we must have $S^B_{ES}(\partial_x \phi_+ , \partial_x \phi_-) = S^A_{ES}(-\partial_x \phi_+, -\partial_x \phi_-)$. Combining these we have $S^A_{ES}(\partial_x \phi_+, \partial_x \phi_-) = S^A_{ES}(-\partial_x \phi_-, -\partial_x \phi_+)$. 

\subsection{Hemisphere swapping symmetry in the effective description of the $\nu = 2/3$ RSES}
To describe the RSES on the sphere, we previously used the hierarchy construction to understand the edge structure at the equator. One can also view the bosonic $\nu = 2/3$ state as an integer quantum Hall state of composite fermions, where the CF particles occupy the two lowest effective Landau levels which are often referred to as $\Lambda$ levels. One may then wonder how we should view the edge structure from this point of view. 

One may expect that the $A$ edge should be composed of two modes where each represents the density of composite fermions in the two $\Lambda$ levels. However, as the $\nu = 2$ case would indicate, we should in fact describe the $A$ edge as the density of composite fermions in the $\psi_{\pm, m}$ orbitals. From the composite fermion perspective we should describe the edge using the density modes $\partial_x \phi_\pm$. Once again, the hemisphere swapping symmetry of our entire wave function gives $S^A_{ES}(\partial_x \phi_+ , \partial_x \phi_-) = S^B_{ES}(\partial_x \phi_-, \partial_x \phi_+)$ and the local conservation of the number of composite fermions in either set of orbitals gives $S^B_{ES}(\partial_x \phi_+ , \partial_x \phi_-) = S^A_{ES}(-\partial_x \phi_+, -\partial_x \phi_-)$. Hence, we still have $S^A_{ES}(\partial_x \phi_+, \partial_x \phi_-) = S^A_{ES}(-\partial_x \phi_-, -\partial_x \phi_+)$ as for the $\nu = 2$ case.

We now just need to determine how the transformation $\partial_x \phi_\pm \rightarrow -\partial_x \phi_\mp$ acts on the charge and neutral modes used in the main text. In the $K$-matrix description of abelian quantum Hall fluids it has been pointed out that the composite fermion theory should correspond to the symmetric basis \cite{Read1990, Hansson, Hansson2017}. In the symmetric basis our $K$-matrix and charge vector, $t$, for this state are,
\begin{equation}
    K = 
    \begin{pmatrix} 
        2 & 1 \\
        1 & 2
    \end{pmatrix} 
    \quad
    t = 
    \begin{pmatrix}
        1 \\
        1 \\
    \end{pmatrix}
\end{equation}
In the hierarchy basis,
\begin{equation} \label{KMatrix}
    K = 
    \begin{pmatrix} 
        2 & -1 \\
        -1 & 2
    \end{pmatrix} 
    \quad
    t = 
    \begin{pmatrix}
        1 \\
        0 \\
    \end{pmatrix}
\end{equation}

We denote the edge modes of the hierarchy basis as $\phi_1$ and $\phi_2$. Note that these do not correspond to the modes $\Tilde{\varphi}^{(1)}$ and $\Tilde{\varphi}^{(2)}$ as the edge action is diagonal for these modes, but there are cross terms in the hierarchy basis. On the requirement that the edge action should be diagonal in $\Tilde{\varphi}^{(1)}$ and $\Tilde{\varphi}^{(2)}$, and that the edge excitation structure should be the same as that detailed in the main text, one can deduce that,
\begin{equation}
    \begin{pmatrix}
        \phi_1 \\
        \phi_2 \\
    \end{pmatrix}
    = 
    \begin{pmatrix}
        1/\sqrt{2} & 1/\sqrt{6} \\
        1/\sqrt{2} & -1/\sqrt{6} \\
    \end{pmatrix}
    \begin{pmatrix}
        \Tilde{\varphi}^{(1)} \\
        \Tilde{\varphi}^{(2)} \\
    \end{pmatrix}
\end{equation}

Further, the $SL(2, \mathbb{Z})$ transformation from the symmetric basis to the hierarchy basis is,
\begin{equation}
    \begin{pmatrix}
        \phi_1 \\
        \phi_2 \\
    \end{pmatrix}
    = 
    \begin{pmatrix}
        1 & 1 \\
        0 & 1 \\
    \end{pmatrix}
    \begin{pmatrix}
        \phi_+ \\
        \phi_- \\
    \end{pmatrix}
\end{equation}

Combining these transformation gives the transformation from the symmetric basis modes to the charged and neutral modes of the main text to be,
\begin{equation}
    \begin{pmatrix}
        \varphi^{(1)} \\
        \varphi^{(2)} \\
    \end{pmatrix}
    = 
    \begin{pmatrix}
        -\sqrt{\frac{3}{2}} & -\sqrt{\frac{3}{2}} \\
        -\frac{1}{\sqrt{2}} & \frac{1}{\sqrt{2}} \\
    \end{pmatrix}
    \begin{pmatrix}
        \phi_+ \\
        \phi_- \\
    \end{pmatrix}
\end{equation}
Hence, we can now interpret the neutral mode as representing the difference in densities of composite fermions in the $\pm$ orbitals at the edge. Thus, we see that if $S^A_{ES}$ is invariant under $\partial_x \phi_\pm \rightarrow -\partial_x \phi_\mp$, then in terms of the charged and neutral modes it must be invariant under $i\partial \varphi^{(1)} \rightarrow -i\partial \varphi^{(1)}$, $i\partial \varphi^{(2)} \rightarrow i\partial \varphi^{(2)}$. 

\subsection{Numerical test of this symmetry}
\begin{figure}
    \centering
    \includegraphics{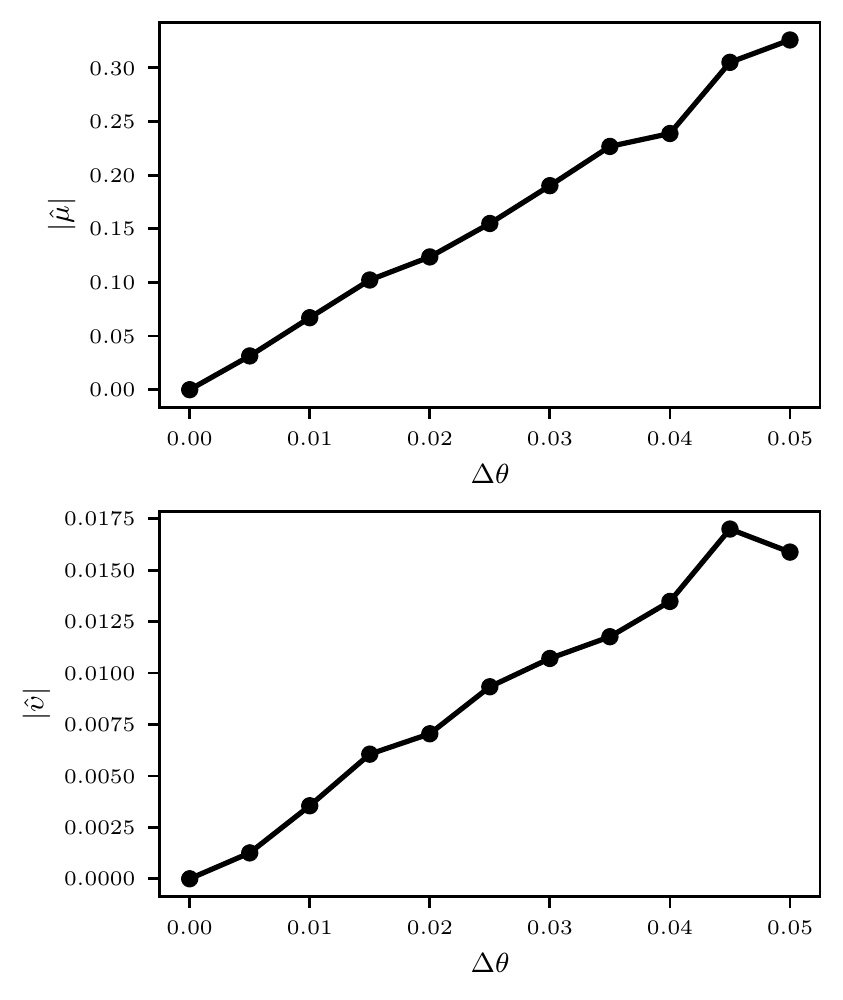}
    \caption{The dependence of the fitted parameters $\hat{\mu}$ and $\hat{\upsilon}$, defined in Eq. \ref{NonSymmetrixEA}, on the shift of the $\theta$ coordinate of the real-space cut from $\pi /2$, $\Delta \theta$. No error bars have been included. At $\Delta \theta = 0$ the entanglement action is expected to have the symmetry proposed in Sec. \ref{Section:23ModelAction}, which requires $\hat{\mu} = \hat{\upsilon} = 0$. One can see that as $\Delta \theta$ goes to zero both $|\hat{\mu}|$ and $|\hat{\upsilon}|$ fall to zero, thus giving some evidence of the proposed symmetry.}
    \label{fig:SymTest}
\end{figure}

Here we shall present a simple numerical test of how this symmetry is manifest in the effective description. In this test the real-space cut is slightly shifted so that it's $\theta$ coordinate (in spherical coordinates) is moved by some amount $\Delta \theta$ (i.e. the real-space cut will be the line defined by $\theta = \pi /2 + \Delta \theta$) (see Fig. \ref{fig:sphere}). We then expect that the coefficients of the terms that are not invariant under the proposed symmetry should now have non-zero values. As we move $\Delta \theta$ to zero these coefficients should fall to zero.

We have only tested this for terms at scaling dimension two in the entanglement action. At scaling dimension two the non-symmetric entanglement action takes the form (not including vertex terms as disscussed in the main text),
\begin{equation} \label{NonSymmetrixEA}
    S^\vee_{ES} = S^{\{ 2 \}}_{ES} +  \oint \frac{dz}{2 \pi i}[ \hat{\mu} i\partial \varphi^{(1)}(z) + \hat{\upsilon} i\partial \varphi^{(1)}(z) i\partial \hat{\varphi}^{(2)}(z) ]
\end{equation}
where $\hat{\mu}$ and $\hat{\upsilon}$ are numerical coefficients to be determined.

We have then fitted this entanglement action to the, numerically calculated, RSES of $N = 38$ bosons in the $\nu = 2/3$ bosonic Jain wave function, at various small values of $\Delta \theta$. The RSES is calculated using the methods detailed in Appendix \ref{Appendix:Numerics} and the entanglement action is fit over all $N_A$ sectors from $16$ to $22$ using the Fit1 procedure detailed in Appendix \ref{Appendix:FittingProcedureAndFittedValues}, where now we have used the full spectrum values $\xi$ rather than the relative levels in each $N_A$ sector so as to be able to fit the parameter $\hat{\mu}$.

The result of this test can be seen in Fig. \ref{fig:SymTest}. One can clearly see that both $\hat{\mu}$ and $\hat{\upsilon}$ drop to zero as $\Delta \theta$ goes to zero. This then gives some evidence that the entanglement action is invariant under $i\partial \varphi^{(1)} \rightarrow -i\partial \varphi^{(1)}$, $i\partial \varphi^{(2)} \rightarrow i\partial \varphi^{(2)}$.

\section{Angular Momentum Calculations} \label{Appendix:AngularMomentum}
Here we shall show how certain quantum numbers in the effective edge theory are related to the angular momentum of the entire $\nu=2/3$ quantum Hall droplet. Further, we shall show that the state with the lowest angular momentum eigenstate of the reduced density matrix in the $\Delta N_A = 0$ sector differs in angular momentum from the state with lowest entanglement pseudo-energy by $O(N)$.

\subsection{Angular momentum in the effective edge theory}
As was stated in Sec. \ref{ConstructionOf23Model}, from the point of view of the standard Halperin-Haldane hierarchy, the $\nu = 2/3$ can be thought of as a $\nu=1/2$ droplet with a $\nu = 1/6$ droplet on top, which is composed of quasi-particles \cite{Haldane1983, Halperin1984}. For the purpose of clarity, we shall work in the disk geometry for the moment.  Following the Halperin-Haldane construction one can write a wave function for this case as,
\begin{equation}
    \begin{split}
        \psi(z_1, \dots, z_N) =& \int \prod_{k = 1}^{N_q} d^2w \prod_{i<j}^{N_q} (w_i - w_j)^2 e^{-\sum_i |w_i|^2/(4l_{B*}^2)} \\
        &\times \prod_{i=1}^{N_q}\prod_{j=1}^N (\bar{w}_i - \partial_{z_j}) \\
        &\times \prod_{i<j}^N (z_i - z_j)^2 e^{-\sum_i |z_i|^2/(4l_B^2)}
    \end{split}
\end{equation}
where $N$ is the number of particles, $N_q$ is the number of quasi-particles, $l_B$ is the magnetic length for the underlying particles and $l_{B*}$ is the effective magnetic length for the quasi-particles. Note that where is the quasi-particle droplet forms is precisely where the $\nu = 1/6$ droplet is located. 

One can easily show, by transforming $z \rightarrow e^{i \theta}z$, that the angular momentum of this wave function is,
\begin{equation}
    M = N(N-1) - N_q N - N_q(N_q - 1)
\end{equation}
This can then be expressed in terms of the $\nu = 1/2$ droplet charge, $Q^{(1)} = N - \frac{N_q}{2}$, and the $\nu = 1/6$ droplet charge, $Q^{(2)} = \frac{N_1}{2}$, as,
\begin{equation} \label{totalAngularMomentum}
    M = Q^{(1)}( Q^{(1)} - 1 ) + 3Q^{(2)}(Q^{(2)} - 1)
\end{equation}
Notice that this is exactly what one would expect if one naively took the formula for angular momentum of a $\nu = 1/2$ and $\nu = 1/6$ droplet separately and added them together (i.e. the same formula one would get for a two layered system where one layer is a $\nu = 1/2$ Laughlin droplet, the other was a $\nu = 1/6$ droplet and the two layers are completely non-interacting). For any state that is adiabatically connected to this state, this formula should still apply. 

If we add charge $\Delta Q^{(1)}$ to the $\nu = 1/2$ edge and $\Delta Q^{(2)}$ to the $\nu = 1/6$ edge, then the angular momentum will change by,
\begin{equation}
    \begin{split}
        \Delta M_{\text{charge}} =& (\Delta Q^{(1)})^2 + 3(\Delta Q^{(2)})^2 \\ 
        &+ 2(Q^{(1)} - 1/2)\Delta Q^{(1)} \\
        &+ 6(Q^{(2)} - 1/2)\Delta Q^{(2)}
    \end{split}
\end{equation}
Written in terms of the zeroth modes of the effective edge theory we have,
\begin{equation}
    \begin{split}
        \Delta M_{\text{charge}} =& \frac{1}{2}(\Tilde{a}^{(1)}_0)^2 + \frac{1}{2}(\Tilde{a}^{(2)}_0)^2 \\ 
        &+ \sqrt{2}(Q^{(1)} - 1/2)\Tilde{a}^{(1)}_0 + \sqrt{6}(Q^{(2)} - 1/2)\Tilde{a}^{(2)}_0
    \end{split}
\end{equation}
Furthermore, it can straightforwardly be seen from Wen's effective edge theory\cite{Wen1995} that the contribution to the angular momentum from the edge phonons is,
\begin{equation}
    \Delta M_{\text{phonons}} = \sum_{n > 0}( \Tilde{a}^{(1)}_{-n}\Tilde{a}^{(1)}_n + \Tilde{a}^{(2)}_{-n}\Tilde{a}^{(2)}_n)
\end{equation}
Thus, the total angular momentum is,
\begin{equation}
    \begin{split}
        \Delta M &= \Delta M_{\text{charge}} + \Delta M_{\text{phonon}} \\
        &= L_0 + \sqrt{2}(Q^{(1)} - 1/2)\Tilde{a}_0^{(1)} + \sqrt{6} (Q^{(2)} - 1/2)\Tilde{a}_0^{(2)}
    \end{split}
\end{equation}
which gives us Eq. \ref{AngularMomentumEq}.

\subsection{Angular momentum difference is $O(N)$ between lowest $M$ eigenstate and lowest pseudo-energy state} \label{AngularMomentumDiffSection}
In Sec. \ref{Section:23ModelAction} it was asserted that the state with the lowest angular momentum eigenstate of the reduced density matrix in the $\Delta N_A = 0$ sector differs in angular momentum from the state with lowest entanglement pseudo-energy by $O(N)$ (assuming, of course, that we are below the entanglement gap and that the cut and glue approach is valid). This is rather important in ensuring $\hat{S}^{\{ 2 \}}_{ES}$ and $\hat{S}^{\{ 3 \}}_{ES}$ can be approximated well with a finite number of terms of low scaling dimension at large but finite system size. We shall demonstrate this difference is $O(N)$ by first showing the difference in angular momentum between the lowest angular momentum state and the average angular momentum is $O(N)$. Then, we will demonstrate that the difference in angular momentum from the lowest pseudo-energy state and the average angular momentum is also $O(N)$. This then implies that the difference in angular momentum between the lowest angular momentum state and the lowest pseudo-energy state is $O(N)$.

From now on we shall be working on the sphere. When we speak of angular momentum on the sphere in this section we really mean $M_{\text{sphere}} + NN_{\phi}/2$, where $N_\phi$ is the number of flux quanta through the sphere, $M_{\text{sphere}}$ is the actual angular momentum on the sphere and $N$ is the number of particles present. This form of angular momentum means one can more straightforwardly transform from the plane geometry to the sphere geometry in such a way that Eq. \ref{totalAngularMomentum} remains valid. We will use the the coordinates $v$ and $z$, where $v$ and $u$ are the usual spinor coordinates (see Appendix \ref{Appendix:Symmetry}) and $z = u/v$. In these coordinates the lowest Landau level orbitals are labelled by an integer $m \in \{0 ,1 , \dots, N_\phi \}$ and take the form $\psi_m = v^{N_\phi}z^m$ (note $m$ is defined differently here than in Appendix \ref{Appendix:Symmetry}). One can then see how $z$ can be used to map to the plane geometry and we shall take the $A$ hemisphere to be in the $|z| < 1$ region of the sphere.

Firstly, let us assume we are in the $N_A = N/2$ sector and that $N/2$ is odd (the following arguments can easily be altered slightly to deal with the even case). From the cut and glue approach, the eigenstate of the reduced density matrix which has the lowest angular momentum below the gap must correspond to some configuration of $\nu = 1/2$ and $\nu = 1/6$ with no edge phonon excitations. From Eq. \ref{totalAngularMomentum} one can calculate that the lowest $M$ configuration is when $Q^{(1)} = (3N_A - 1)/4$ and $Q^{(2)} = (N_A + 1)/4$, with the total angular momentum being $3(N_A-1)^2/4$.

One can also compute the average angular momentum in $A$ when the full wave function is placed on the sphere. Applying the method given by Girvin et al. \cite{Girvin1987}, one can show the single particle reduced density matrix on the sphere of a rotationally symmetric wave function on the sphere is,
\begin{equation}
    \rho(x_1, x_2) = \rho_0 \bar{v}_1^{N_\phi} v_2^{N_\phi} (1 + \bar{z}_1z_2)^{N_\phi} 
\end{equation}
where $x$ is a shorthand for a position on the sphere and $\rho_0$ is the particle density. 

One can then compute the angular momentum density,
\begin{equation}
\begin{split}
    M(x) &= z\partial_{z_2}\rho(x,x) \\
    &= N_\phi\rho_0 \frac{|z|^2}{1 + |z|^2}
\end{split}
\end{equation}
Now let $R$ be the radius of the sphere and $\mathcal{S} = \nu^{-1}N - N_\phi$ be the shift. We can then find the average angular momentum in $A$, $\braket{M_A}$,
\begin{equation}
    \begin{split}
        \braket{M_A} =& \int_A R^2 d\Omega M(x) \\
        =& 2\pi R^2 N_\phi \rho_0 \int_{\frac{\pi}{2}}^\pi d \theta \sin (\theta) \frac{\cot^2 (\theta/2)}{1 + \cot^2(\theta/2)} \\
        &= \frac{N_\phi N}{8} \\
        &= \frac{\nu^{-1} N^2 - \mathcal{S}N}{8} \\
        &= \frac{3}{4}N_A^2 + O(N)
    \end{split}
\end{equation}

Thus, we can see the difference between the average angular momentum in $A$ and the minimum angular momentum eigenstate of the reduced density matrix in the $N_A = N/2$ sector is $O(N)$.

Now we move over to our effective description. The angular momentum relative to the lowest pseudo energy state with charges $Q^{(1)}$ and $Q^{(2)}$ in the $\nu = 1/2$ and $\nu = 1/6$ droplets respectively is given by,
\begin{equation}
    \Delta M = L_0 - \sqrt{\frac{3}{2}}(N/2 - 1/2)a^{(1)}_0 + \frac{1}{\sqrt{2}}( Q^{(1)} - 3Q^{(2)} + 1 )a^{(2)}_0
\end{equation}
If we let $\zeta = \frac{1}{\sqrt{2}}( Q^{(1)} - 3Q^{(2)} + 1 )$, then one can see that the minimum angular momentum in the $N_A = N/2$ sector relative to the minimum pseudo energy state is $\Delta M_{\text{min}} = -\frac{\zeta^2}{2}$. Note that as we are defining our effective theory around the minimum pseudo-energy state in general we should expect $\zeta$ to depend on $N$.

We would now like to compute the average relative angular momentum to this state in $A$ using our effective description, in the large $N$ limit. In the large $N$ limit, we can approximate $S_{ES}$ with $S^{\{ 2 \}}_{ES}$ and thus our partition function can be written as,
\begin{equation}
    \begin{split}
        Z =& \text{Tr}e^{-S^{\{2\}}_{ES}} \\
        =& \bigg ( \prod_{n=1}^\infty \frac{1}{1 - e^{-\beta n}} \bigg )
        \bigg ( \prod_{m=1}^\infty \frac{1}{1 - e^{-\gamma m}} \bigg ) \\ 
        &\times \sum_{q \in \Gamma} e^{-\frac{\beta}{2}q_1^2 - \frac{\gamma}{2}q_2^2 - \alpha q_2} \\
        =& \frac{e^{-\frac{\beta + \gamma}{24}}}{\eta\big ( \frac{i\beta}{2\pi} \big ) \eta\big ( \frac{i\gamma}{2\pi} \big )} \sum_{q \in \Gamma} e^{-\frac{\beta}{2}q_1^2 - \frac{\gamma}{2}q_2^2 - \alpha q_2}
    \end{split}
\end{equation}
where $\eta(\tau)$ is the Dedekind eta-function. Note that all $\alpha$, $\beta$ and $\gamma$ scale as $1/\sqrt{N}$. Thus, it may not seem obvious how to proceed to work in the large $N$ limit with our partition function written in this form. One can proceed with the calculations below by converting the sum into an integral. However, we will take a more controlled approach here. Using Poisson resummation (see section 10.4.2 in Di Francesco et al. \cite{DiFrancesco1997}) and that $\eta(-1/\tau) = \sqrt{-i\tau}\eta(\tau)$, we get,
\begin{equation}
    \begin{split}
        Z =& \frac{e^{-\frac{\beta + \gamma}{24}}}{\eta\big ( \frac{2\pi i}{\beta} \big ) \eta\big ( \frac{2\pi i}{\gamma} \big )\text{Vol}(\Gamma)} \sum_{p \in \Gamma^*} e^{-\frac{2\pi^2}{\beta}p_1^2 - \frac{2\pi^2}{\gamma}(p_2 + \frac{\alpha}{2\pi i})^2} \\
        =& e^{-\frac{\beta + \gamma}{24} + \frac{\pi^2}{6}(1/\beta + 1/\gamma)} \\
        &\times \bigg ( \prod_{n=1}^\infty \frac{1}{1 - e^{- \frac{4\pi^2}{\beta} n}} \bigg )
        \bigg ( \prod_{m=1}^\infty \frac{1}{1 - e^{-\frac{4\pi^2}{\gamma} m}} \bigg ) \\
        &\times \text{Vol}(\Gamma)^{-1} \sum_{p \in \Gamma^*} e^{-\frac{2\pi^2}{\beta}p_1^2 - \frac{2\pi^2}{\gamma}(p_2 + \frac{\alpha}{2\pi i})^2}
    \end{split}
\end{equation}
where Vol$(\Gamma)$ is the unit cell volume of $\Gamma$.

Noting that $\alpha, \beta, \gamma \propto 1/\sqrt{N}$ we must have that at large $N$,
\begin{equation}\label{logZ}
    \begin{split}
        \ln Z =& \frac{\pi^2}{6} \bigg (\frac{1}{\beta} + \frac{1}{\gamma} \bigg ) + \frac{\alpha^2}{2\gamma} - \frac{\beta + \gamma}{24} \\
        &+ \ln \text{Vol}(\Gamma) + \text{subleading corrections} 
    \end{split}
\end{equation}

To calculate the average relative angular momentum we can break it down into calculating $\braket{L_0}$, $\braket{a^{(1)}_0}$ and $\braket{a^{(2)}_0}$. Note that $\braket{a^{(1)}_0} = 0$ by symmetry and so we only need to compute $\braket{L_0}$ and $\braket{a^{(2)}_0}$ in the large $N$ limit.

To compute $\braket{L_0}$ one can define a $Z(\omega)$ by transforming $\beta \rightarrow \beta - \omega$ and $\gamma \rightarrow \gamma - \omega$. Then we have $\braket{L_0} = \frac{\partial\ln Z(0)}{\partial\omega}$. In the large $N$ limit this gives,
\begin{equation}
    \braket{L_0} = \frac{\pi^2}{6} \bigg (\frac{1}{\beta^2} + \frac{1}{\gamma^2} \bigg ) + \text{subleading corrections}
\end{equation}
It then follows that $\braket{L_0} = O(N)$.

Next, $\braket{a^{(2)}_0} = -\frac{\partial\ln Z}{\partial\alpha}$, and so we have, again for large $N$,
\begin{equation}
    \braket{a^{(2)}_0} = -\frac{\alpha}{\gamma}
\end{equation}
which is a constant for large $N$.

This then gives,
\begin{equation}
    \begin{split}
        \braket{\Delta M} =&  \frac{\pi^2}{6} \bigg (\frac{1}{\beta^2} + \frac{1}{\gamma^2} \bigg ) - \zeta \frac{\alpha}{\gamma} \\
        &+ \text{subleading corrections}
    \end{split}
\end{equation}
Since $Q^{(1)} + Q^{(2)} = N$ and $Q^{(j)} \geq 0$ we must have that $\zeta$ is at most $O(N)$ (i.e. $\zeta$ cannot grow any faster than $N$). It then simply follows that $\braket{\Delta M} = O(N)$ in the large $N$ limit. 

Putting this altogether, we have that the difference in angular momentum from the eigenstate of the reduced density matrix, in the $N_A = N/2$ sector, with the lowest angular momentum and the average angular momentum is $O(N)$ and the difference in angular momentum from the state with lowest entanglement pseudo-energy and the average angular momentum is also $O(N)$. Hence, the difference in angular momentum from the lowest pseudo-energy state and the eigenstate of the reduced density matrix, in the $N_A = N/2$ sector, with the lowest angular momentum is $O(N)$. QED.

\section{Numerical computation of RSES} \label{Appendix:Numerics}
\begin{figure}
\includegraphics[width=0.7\columnwidth]{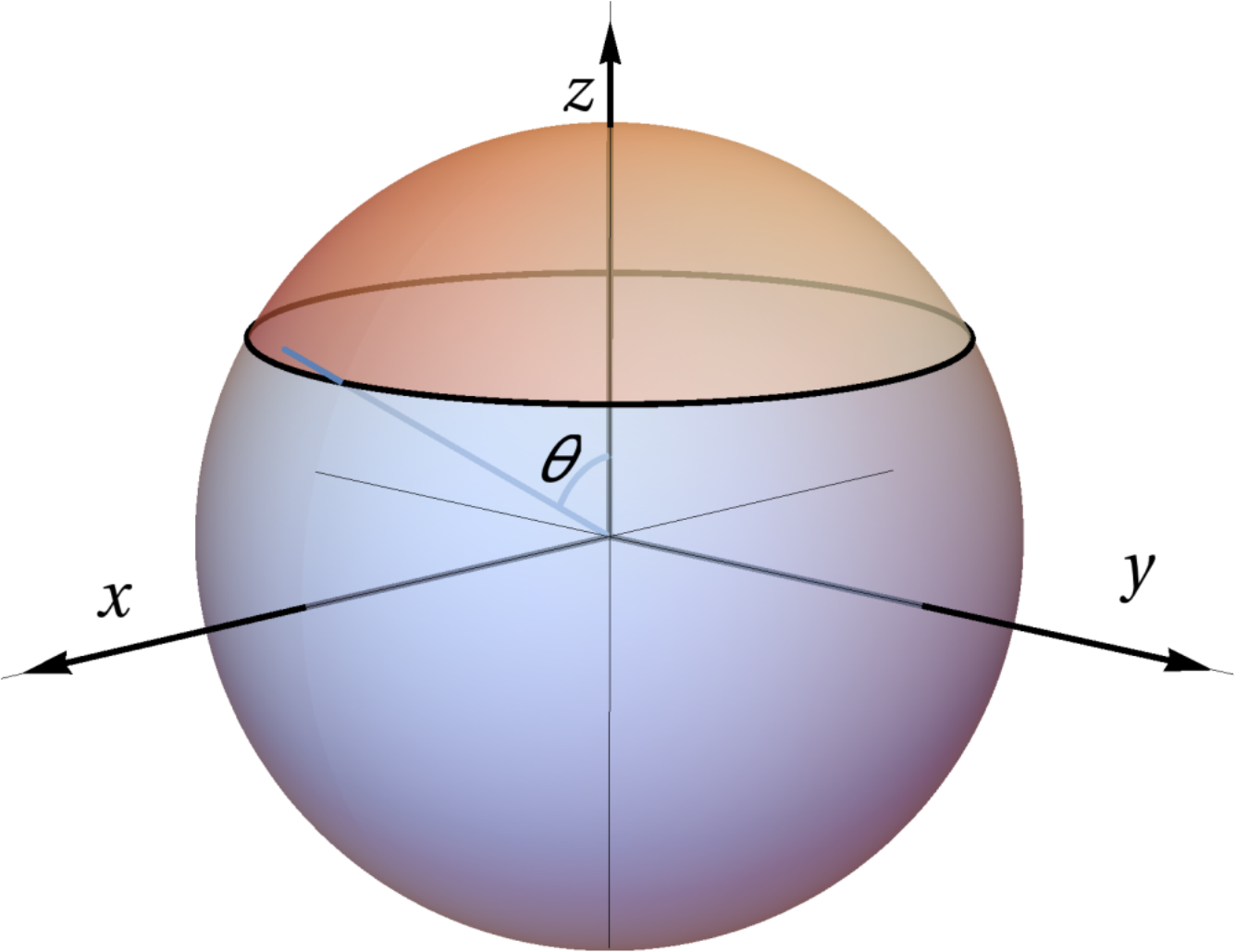}
\caption{Haldane sphere showing the real space cut and the two partitions.\label{fig:sphere}}
\end{figure}
The numerical calculations of the real space entanglement spectra shown in this work use the ideas introduced in Ref. \cite{Rodriguez2013}.  To enable computations on larger systems, the techniques here are implemented using higher precision arithmetic and extended to allow computation of spectra across particle number sectors. In this section, we summarize the technique and sketch details of implementation. Calculations were performed in the spherical geometry\cite{Haldane1983}, but we will represent the particle coordinates as $z$ for convenience.

The bosonic composite fermion state at filling fraction $2/3$, whose real space entanglement spectrum is analyzed in this work can be written as 
\begin{equation}
\psi_{\frac{2}{3}} = \mathcal{P}_{\rm LLL} \phi_1 \phi_2
\end{equation}
where $\phi_1$ is the fermionic wave function of $N$ particles fully filling a Landau level in $N-1$ flux quanta through the sphere, and $\phi_2$ is the fermionic wave function with two filled Landau levels in $N/2-2$ flux quanta. The operator $\mathcal{P}_{\rm LLL}$ projects the wave function into the lowest Landau level. 

To compute the real space entanglement we treat the particles as distinguishable and partition them into two sets -- $Z_A$ containing $z_1,z_2\dots z_{N_A}$ and $Z_B$ containing the remaining $N_B=N-N_A$ particles. To separate out these coordinates inside $\psi_{\frac{2}{3}}$, we note that the Slater determinant structure of $\phi_1$ and $\phi_2$ allows us to write the wave function as
\begin{equation}
\psi_{\frac{2}{3}}=\mathcal{P}_{\rm LLL} \sum_{\sigma,\tau\in {\rm perm}(N)} (-1)^{\sigma\tau} \prod_{i=1}^{N} \; \eta_{m_{\sigma(i)}}(z_i) \; \eta_{s_{\tau(i)}}(z_i).\label{eq:permutationform}
\end{equation}
Here $\eta_k$ is the single particle momentum eigenstate labeled by a pair $k\equiv(n,l)$ that combines the Landau level index $n$ and the azimuthal angular momentum quantum number $l$. In the above expression $m_i\in M$ and $s_i\in S$ where $M$ and $S$ are the sets of occupied single particle states $(n,l)$ in the Slater determinant forms of $\phi_2$  and $\phi_1$. We associate an ordering within $M$ and $S$ to keep track of signs when the particles are permuted.

We can sum over the permutations within each set $Z_A$ and $Z_B$ to write the function as  
\begin{equation}
\psi_{\frac{2}{3}}=\mathcal{P}_{\rm LLL}\sum_\lambda {\rm sign}(\lambda) \xi^A_\lambda (Z_A) \xi^B_\lambda (Z_B)\label{eq:entWfForm}
\end{equation}
Here $\lambda$ sums over the $(\frac{N!}{N_A!N_B!})^2$ ways to divide $M$ into two parts $(M_A,M_B)$ and $S$ into two parts $(S_A,S_B)$ of sizes $(N_A,N_B)$. 
The functions $\xi^A,\xi^B$ indexed by $\lambda$, called the entanglement wave functions are given by the 
\begin{gather}
\xi_A=\mathcal{P}_{LLL} D_{M_A}(Z_A)D_{S_A}(Z_A)\nonumber\\
\xi_B=\mathcal{P}_{LLL} D_{M_B}(Z_B)D_{S_B}(Z_B)\label{eq:xiDefinition}
\end{gather} 
where $D_X(Y)$ is the Slater determinant state in which particles in $Y$ occupy the orbitals in $X$. The sign of $\lambda$ is the product of the signs association with the permutations that take $M$ to the merged set $(M_A,M_B)$ and $S$ to the merged set $(S_A,S_B)$. For instance if $S=(k_1,k_2,k_3)$ and $S_A=(k_1,k_3)$, $S_B=k_2$, the sign of the permutation that takes $S$ to $(S_A,S_B)=(k_1,k_3,k_2)$ is $-1$.

We introduce a real-space cut partitioning the sphere as shown in Fig \ref{fig:sphere}. 
and then project the wave function into a subspace where the particles in sets $Z_A$ and $Z_B$ are inside the regions $A$ and $B$ respectively. The projected function is the same as Eq. \ref{eq:permutationform} if all particles $Z_A$ and $Z_B$ are in $A$ and $B$ respectively; and is zero otherwise. 
The entanglement spectrum can now be computed by diagonalizing the reduced density matrix $\rho_A$ of this projected wave function. Thanks to the azimuthal rotational symmetry of the wave function and the cut, $\rho_A$ is block diagonal with blocks labeled by azimuthal angular momentum. The block associated with angular momentum $L_A$ can be constructed by retaining only those terms of the summation in Eq. \ref{eq:entWfForm} in which the entanglement function $\xi^A$ has momentum $L_A$.

Since the angular momentum operator commutes with the lowest Landau level projection and also with the projection into the real space partitions, the momentum of $\xi^{A/B}$ is the same as the combined momentum of $M_{A/B}$ and $S_{A/B}$. Therefore the entanglement functions $\xi^A$ with angular momentum $L_A$ can be enumerated by enumerating all size $N_A$ subsets of $M$ and $S$ with total momentum $L_A$. The complementary subset of $M$ and $S$ define the unique state $\xi^B$ complementary to each each $\xi^A$. 

The reduced density matrix restricted to this momentum sector can be written as 
\begin{equation}
\rho_A = \sum_{\lambda,\lambda'} \left |  \xi^A_\lambda \right \rangle \left \langle \xi^B_{\lambda'} | \xi^B_{\lambda} \right \rangle \left \langle \xi^A_{\lambda'}\right |\nonumber
\end{equation}
The eigenvalues of this density operator are identical to that of (See Ref. \cite{Rodriguez2013} for details)
\begin{equation}
M_{\nu\mu} = \sum_{\lambda'}\left\langle \xi_{\lambda'}^{A}|\xi_{\mu}^{A}\right\rangle \left\langle \xi_{\lambda'}^{B}|\xi_{\nu}^{B}\right\rangle 
\end{equation}

Note that $\xi^{A/B}$ are functions projected into the respective real space sectors and have support only in $A/B$, therefore the overlaps between entanglement wavefunctions in the above expressions involve integrals within $A/B$ regions only.
Up to a multiplicative constant, the overlaps integrals can be estimated using standard Metropolis Monte Carlo methods.
We chose the sampling distribution to be the square of an entanglement wave function $\xi_0^A$ from the sector with the smallest $L_A$ to estimate the overlap ratio $\left \langle \xi_{\lambda}^{A}|\xi_{\mu}^{A}\right\rangle/\left \langle \xi_{0}^{A}|\xi_{0}^{A}\right\rangle$. 
Similarly, we chose the square of the entanglement wave function $\xi_0^B$ complementary to $\xi_0^A$ as the sampling function to compute the overlap ratio $\left \langle \xi_{\lambda}^{B}|\xi_{\mu}^{B}\right\rangle/\left \langle \xi_{0}^{B}|\xi_{0}^{B}\right\rangle$. Putting them together we can get $M$ up to a multiplicative factor of $|\xi_0^A\xi_0^B|^{-2}$, and accordingly all entanglement energies in an $L_A$ sector can be computed up to a common additive constant. Since the same sampling function is used in all $L_A$ sectors, the entanglement spectrum in all $L_A$ sectors can be computed up to an additive constant. 

Since a different $\xi_0^{A/B}$ was used for calculations in each $N_A$, different $N_A$ sectors were shifted by different additive constant. In order to correctly estimate the relative shifts we separately computed the ratios of $|\xi_0^A\xi_0^B|^2$ between different sectors. This allows estimation of entanglement spectra across multiple $N_A$ sectors upto a additive constant.

The above method does not make any approximations in the calculation. However while evaluating $\xi^{A/B}$, the projection into the lowest Landau level in Eq. \ref{eq:xiDefinition} can only be performed in an approximate way in a manner similar to the Jain-Kamilla projection \cite{Jain1997}. This is implemented by replacing factors of $\bar{z}_i$ in $\xi^{A/B}$ with $\sum_{k=1,k\neq i}^{N_{A/B}}\frac{2}{z_i-z_k}$.

\section{Fitting Procedure and Fitted Parameters} \label{Appendix:FittingProcedureAndFittedValues}
\begin{table}
    \centering
    \begin{tabular}{c c c c c}
        \hline
        \hline
        Parameter & Value(Fit1) & Error & Value(Fit2) & Error \\
        \hline
        $\hat{\alpha}$ & $5.514$ & $6.2 \times 10^{-4}$ & $5.531$ & $2.4 \times 10^{-4}$\\
        $\sqrt{N}\hat{\beta}$ & $2.024$ & $3.1 \times 10^{-3}$ & $2.077$ & $3.0 \times 10^{-3}$\\
        $\sqrt{N}\hat{\gamma}$ & $10.22$ & $1.7 \times 10^{-3}$ & $10.29$ & $1.5 \times 10^{-3}$\\
        \hline
        \hline
    \end{tabular}
    \caption{Fitted parameters for the $S^{\{ 2 \}}_{ES}$ model (Eq. \ref{d2S}) of the RSES of $N = 58$ bosons on the sphere in the $\nu = 2/3$ Jain wave function with the real space cut along the equator. The fitting procedure is detailed in Appendix \ref{Appendix:FittingProcedureAndFittedValues} and the resulting spectra for Fit1 can be seen in Fig. \ref{fig:fittingTest}.  Fit1 and Fit2 include the 4 and 5 lowest angular momentum sectors in each $N_A$ respectively.}
    \label{tab:threeParamFit}
\end{table}

\begin{table}
    \centering
    \begin{tabular}{c c c c c}
        \hline
        \hline
        Parameter & Value(Fit1) & Error & Value(Fit2) & Error\\
        \hline
        $\hat{\alpha}$ & $5.472$ & $6.6 \times 10^{-4}$ & $-5.471$ & $6.0 \times 10^{-4}$\\
        $\sqrt{N}\hat{\beta}$ & $2.039$ & $3.4 \times 10^{-3}$ & $2.100$ & $2.7 \times 10^{-3}$\\
        $\sqrt{N}\hat{\gamma}$ & $10.30$ & $1.7 \times 10^{-3}$ & $10.36$ & $1.4 \times 10^{-3}$\\
        $N\hat{\delta}$ & $0.6461$ & $2.8 \times 10^{-3}$ & $-0.6502$ & $2.0 \times 10^{-3}$\\
        $N\hat{\epsilon}$ & $-0.3113$ & $4.4 \times 10^{-3}$ & $0.3118$ & $4.0 \times 10^{-3}$\\
        \hline
        \hline
    \end{tabular}
    \caption{Fitted parameters for the $S^{\{ 3 \}}_{ES}$ model (Eq. \ref{d3S}) of the RSES of $N = 58$ bosons on the sphere in the $\nu = 2/3$ Jain wave function with the real space cut along the equator. The fitting procedure is detailed in Appendix \ref{Appendix:FittingProcedureAndFittedValues} and the resulting spectra for Fit1 can be seen in Fig. \ref{fig:fittingTest}. Fit1 and Fit2 include the 4 and 5 lowest angular momentum sectors in each $N_A$ respectively.}
    \label{tab:FiveParamFit}
\end{table}
Here we shall detail the fitting procedure used to fit the two models in the numerical tests of Sec. \ref{Section:NumericalTests} and the final fitted parameters for the $N = 58$ test. 

Given some numerically calculated spectrum, $\xi^{(i)}_{RSES}$ we fit any given model spectrum, $\xi^{(i)}_{\text{Model}}$, by minimising,
\begin{equation}
    R = \sum_i \frac{(\Delta\xi^{(i)}_{RSES} - \Delta\xi^{(i)}_{\text{Model}})^2}{2\sigma_i^2} e^{-\frac{(N_A^{(i)} - N/2)^2}{8}}
\end{equation}
where $\Delta \xi$ is the entanglement pseudo-energy relative to the lowest level of the lowest angular momentum sector for the given $N_A$ sector, $N^{(i)}_A$ is the number of particles in spacial region $A$ for $\xi^{(i)}$, $N$ is the total number of particles in the full wave function and $\sigma_i$ is the error in $\xi^{(i)}_{RSES}$. The index $i$ is assigned by first organising the entanglement levels according to their angular momentum and $N_A$ quantum numbers, $(L_z^A, N_A)$, and then ordering the levels in each $(L_z^A, N_A)$ sector from lowest to highest. For all data presented in the main text, the parameters have been fitted using the entanglement levels in the four lowest angular momentum sectors for each $N_A$ sector involved. We will call this procedure Fit1. In this Appendix we will also refer to another fitting procedure which includes the 5 lowest angular momentum sectors in each $N_A$ sector, which we will call Fit2.

Including the errors in the numerical pseudo-energies, $\sigma_i$, allows $R$ to be interpreted as a log likelihood function. The factor $e^{-\frac{(N_A^{(i)} - N/2)^2}{8}}$ is used to control how much the $N_A$ sectors away from $N/2$ are included in the fitting procedure. We have used $\Delta \xi$ rather than just $\xi$ here as the relative pseudo-energy between two different $N_A$ sectors often has a much larger error than the individual numerically calculated levels in a given sector. Thus, it is more optimal to use the relative levels within each $N_A$ sector.

For a given set of model parameters, the spectrum of $S^{\{ j \}}_{ES}$ ($j = 2,3$) is calculated numerically by first computing it's matrix form relative to the normalised version of the basis $\big (\prod_k \prod_{n_k}a^{(k)}_{-n_k} \big )e^{i\sum_l a_l \varphi^{(l)}_0}\ket{0}$, and then using the numpy function numpy.linalg.eigvalsh to calculate the eigenvalues of the resulting matrices. We then minimise $R$ by running the scipy function scipy.optimise.minimise (in default settings) several times and taking the most optimal case. The errors of the parameters are then taken as the square roots of the diagonal elements of the inverse hessian matrix outputted by the scipy.optimise.minimise function. 

For the specific case shown in Fig. \ref{fig:fittingTest} full estimates of the errors of the numerical levels, $\sigma_i$, have been made and the models have been fitted over the $N_A = 29, 28, 27, 26, 25$ using the Fit1 procedure. The fitted values can be seen in Tables \ref{tab:threeParamFit} and \ref{tab:FiveParamFit}. Both tables also include the fitted values using the Fit2 procedure. We have exhibited the values resulting from Fit2 to demonstrate the our fitted parameters have some robustness to the choice of angular momentum sectors, but also to show that the estimated errors are an underestimate. In short there is a systematic error associated with the choice of angular momentum sectors and this has not been included in our estimates. Given this and in the interest of simplicity, the parameters in Figs. \ref{fig:scale2Plot} and \ref{fig:scale3Plot} have been fitted using numerical data where $\sigma_i$ has been crudely estimated to be $\sigma \sim 0.1$ for \textit{all} cases. Furthermore, the parameters in Figs. \ref{fig:scale2Plot} and \ref{fig:scale3Plot} have been fitted over the $N_A = N/2, N/2 - 1, \dots, N/2 - 4$ sectors using the Fit1 sectors.

\section{Further Structure of the Abelian Edge Theories and the Uniqueness of the Boundary State} \label{Appendix:BoundaryState}
In Sec. \ref{sectionOverlaps} it was asserted that the boundary state $\ket{G_*}$ is unique up to a multiplicative constant. We shall now demonstrate this. This is a standard result in boundary conformal field theory. However, for completeness we have included it here.

To properly demonstrate this we first need to discuss the structure of the minimal edge theories of these abelian states in more detail (minimal in the sense of no edge reconstruction and not in the sense of a CFT minimal model). As discussed in the main text, in general our minimal chiral edge theory will contain some integer number, $p$, of chiral boson fields. These fields have the following mode expansion (in radial quantisation coordinates),
\begin{equation}
    \varphi^{(j)} = \varphi^{(j)}_0 - ia^{(j)}_0 \ln z + i\sum_{n \neq 0} \frac{1}{n} a^{(j)}_n z^{-n}
\end{equation}
where,
\begin{equation}
    [a^{(i)}_n, a^{(j)}_m] = n \delta_{n+m, 0}\delta_{ij} \quad \quad [\varphi^{(j)}, a^{(k)}_0] = i \delta_{jk}
\end{equation}
and all other commutation relations are trivial. The vacuum of the theory has the property that $a^{(i)}_n\ket{0} = 0$ for all $n \geq 0$ and $1 \leq i \leq p$. We will use the following shorthand notation, $b \cdot \varphi \equiv \sum_{i=1}^p b_i \varphi^{(i)}$. The Hilbert space of the edge has a basis that is labelled by a lattice vector, $b \in \Gamma$, and $p$ partitions $\lambda^{(i)}$, which is given by,
\begin{equation}
    \ket{b, \lambda^{(i)}} = e^{ib\cdot \varphi_0} \prod_{i=1}^p \prod_{n \in \lambda^{(i)}} a^{(i)}_{-n} \ket{0}
\end{equation}

Suppose we have a state, $\ket{\psi}$, which has the property $a^{(i)}_n\ket{\psi} = 0$ for all $n \geq 0$ and $1 \leq i \leq p$. From the basis given above it follows that such a state must be a complex number times the vacuum state, $\ket{\psi} = \alpha \ket{0}$ with $\alpha \in \mathbb{C}$. This follows in two steps. First, from the property $\ket{\psi}$ satisfies it follows directly that $\ket{\psi}$ must be orthogonal to any basis element involving any $a^{(i)}_{-n}$ operators. Further, as $a^{(i)}_0 \ket{\psi} = 0$, we have $\bra{0}e^{-ib\cdot \varphi_0}a^{(i)}_0\ket{\psi} = b_i \bra{0}e^{-ib\cdot \varphi_0}\ket{\psi} = 0$, which implies for $b \neq 0$, $\bra{0}e^{-ib\cdot \varphi_0}\ket{\psi} = 0$. Thus, this implies that $\ket{\psi} = \alpha \ket{0}$ for $\alpha \in \mathbb{C}$.

Now suppose there is an operator $B$ which commutes with all modes of the fields. From the basis given above, it follows that such an operator is completely defined by its application on the vacuum $B\ket{0}$. As $B$ commutes with all $a^{(i)}_n$ it follows that $a^{(i)}_n B\ket{0} = 0$ for all $n \geq 0$ and $1 \leq i \leq p$. Hence, $B\ket{0} = \alpha\ket{0}$ with $\alpha \in \mathbb{C}$. So $B$ must be some number times the identity, $B = \alpha \mathbb{I}$. Thus, there cannot exist any non-trivial invariant subspaces of the modes of the fields, as for any such subspace the projection operator onto such a subspace must commute with all the modes and is therefore some complex number times the identity operator.

Putting this altogether, we have that the Hilbert space of the edge theory must form an irreducible representation of the chiral algebra formed by all local fields of the theory.

Now we will show that the modes of the vertex operators defined in the main text can generate the entire Hilbert space by application to the vacuum state. By state operator correspondence this implies that all local fields of the theory can be generated through repeated operator product expansion of the vertex operators. 

First by using $\braket{\varphi^{(i)}(z)\varphi^{(j)}(w)} = - \delta_{ij}\ln{(z - w)}$ along with Wick's theorem, we have,
\begin{equation}
    \begin{split}
        V_{\epsilon^j}(z)V_{-\epsilon^j}(w) =& \frac{1}{(z - w)^{\epsilon^i \cdot \epsilon^j}}e^{i\epsilon^j \cdot (\varphi(z) - \varphi(w))} \\
        =& \frac{1}{(z - w)^{\epsilon^j \cdot \epsilon^j}} + \frac{i \epsilon^j \cdot \partial \varphi(w)}{(z - w)^{\epsilon^j \cdot \epsilon^j - 1}} + \dots
    \end{split}
\end{equation}
Thus we can also write,
\begin{equation}
    i \epsilon^j \cdot \partial \varphi(w) = \frac{1}{2 \pi i} \oint_w dz (z-w)^{\epsilon^j \cdot \epsilon^j} V_{\epsilon^j}(z)V_{-\epsilon^j}(w)
\end{equation}
where the integration contour is a small circle around $w$. To make notation lighter we will define $k \equiv \epsilon^j \cdot \epsilon^j$. We can deform the integration contour, so that it is split into a counter-clockwise contour around the origin containing $w$, and an clockwise contour around the origin not containing the point $w$. Further, we can also expand the $(z-w)^k$ factor in the integrand and we then have,
\begin{equation}
    \begin{split}
        &i \epsilon^j \cdot \partial \varphi(w) \\
        =& \frac{1}{2\pi i} \sum_{n = 0}^k (-1)^{k - n} {k \choose n} \oint_{|z| > |w|} dz z^n  w^{k-n} V_{\epsilon^j}(z)V_{-\epsilon^j}(w) \\
        &- \oint_{|z| < |w|} dz z^n w^{k-n} V_{-\epsilon^j}(w)V_{\epsilon^j}(z) \\
        &= \sum_{n = 0}^k (-1)^{k - n} {k \choose n} w^{k - n} [  V_{-\epsilon^j}(w) , V_{\epsilon^j, n - h_j + 1}]
    \end{split}
\end{equation}
where $h_j$ is the scaling dimension of the vertex operator. We can thus write,
\begin{equation}
    \begin{split}
        &\sum_{l = 1}^p \epsilon^j_l a^{(l)}_m \\ 
        &= \sum_{n = 0}^k (-1)^{k - n} {k \choose n} [  V_{-\epsilon^j, k + m - n - h_j + 1} , V_{\epsilon^j, n - h_j + 1}]
    \end{split}
\end{equation}
Note that in the last line one may need consider anti-commutation relations depending on the conformal spin of the vertex operators involved. 
As the lattice vectors $\epsilon^j$ are linearly independent we can take linear combinations of our expression for $\sum_{l = 1}^p \epsilon^j_l a^{(l)}_m$, with different $j$, to obtain an expression for $a^{(l)}_m$. Hence, the $a^{(l)}_m$ operators can be expressed as linear combinations of commutators of the vertex operators. 

Now, we also have that the operator product expansion between any generic vertex operators is, $V_a(z)V_b(w) = (z-w)^{a \cdot b}V_{a + b}(w) + \dots$. Hence, we can apply the same calculation from above to inductively express the modes of any vertex operator in terms of the modes of $V_{\epsilon^j}$. Also since $V_b(0)\ket{0} = V_{b, -h_b}\ket{0}$, we can also write any $e^{i b \cdots \varphi_0}\ket{0}$ as some combination of the modes of $V_{\epsilon^j}(z)$ acting on the vacuum. We have thus demonstrated that the basis given in this section can be generated by modes of the vertex operators, $V_{\epsilon^j}(z)$, applied on the vacuum state. Hence, any state in the Hilbert space of this theory can be generated by repeated application of the modes of the vertex operators on the vacuum state.

Now we move on to showing the uniqueness of our boundary state. Recall that $\ket{G_*}$ must satisfy Eq. \ref{boundaryCon}. Now define an anti-unitary operator $U$ that maps from the Hilbert space of the $A$ edge to the Hilbert space of the $B$ edge, such that $U\ket{0} = \overline{\ket{0}}$ (i.e. it maps the vacuum of one edge to the vacuum of the other) and $U V_{\epsilon^i, n} = \overline{V}_{\epsilon^i, n} U$. Note this completely defines $U$ as both the $A$ and the $B$ edges form irreducible representation of the chiral algebra generated by the vertex operators. We can write $\ket{G_*} = \sum_{ij} G_{ij} \ket{j} \otimes (U\ket{i}) \equiv \sum_i G \ket{i} \otimes (U\ket{i})$, where $G$ is an operator that only acts on the Hilbert space of the $A$ edge. Our boundary condition now gives $\sum_i ( V_{\epsilon^l, n} G \ket{i} ) \otimes (U\ket{i}) - (G \ket{i}) \otimes (U V_{-\epsilon^l, -n} \ket{i} ) = 0$. Then we have,
\begin{equation}
\begin{split}
    &\bra{j} \otimes \bra{U k} \sum_i \big [ ( V_{\epsilon^l, n} G \ket{i} ) \otimes (U\ket{i}) \\
    &- (G \ket{i}) \otimes (U V_{-\epsilon^l, -n} \ket{i} ) \big ] \\
    & = \sum_i \bra{j} V_{\epsilon^l, n} G\ket{i} \braket{i | k} - \bra{j}G\ket{i} \bra{i}V_{\epsilon^l, n} \ket{k} \\
    &=  \bra{j}V_{\epsilon^l, n} G \ket{k} - \bra{j} G V_{\epsilon^l, n} \ket{k} \\
    &= \bra{j} [V_{\epsilon^l, n}, G]\ket{k} \\
    &= 0
\end{split}
\end{equation}
So we have $ [V_{\epsilon^l, n}, G] = 0 $. As the $A$ edge forms an irreducible representation of the chiral algebra generated by the vertex operators, by Schur's lemma we then have that $G$ is a multiple of the identity. One can also use the fact that if $G$ commutes with these vertex operators, it must commute with all the modes of the chiral boson fields and thus is some complex number times the identity by the earlier discussion of this section. We know $\ket{G_*} \neq 0$ so we can choose the normalization of $|G_*\rangle$ such that,
\begin{equation}
    \ket{G_*} = \sum_i \ket{i} \otimes (U \ket{i})
\end{equation}

Strictly speaking, in the main text when we identified an $A$ edge state $\ket{a}$ with a $B$ edge state $\overline{\ket{a}}$ we mean $U\ket{a} \equiv \overline{\ket{a}}$. Hence, it follows,
\begin{equation}
    \begin{split}
        \bra{a}\overline{\bra{b}}\ket{G_*} &= \bra{a}\bra{Ub}\ket{G_*} \\
        &= \bra{a}\bra{Ub} \sum_i \ket{i} \otimes (U \ket{i}) \\
        &= \sum_i \braket{a | i}\braket{i | b} \\
        &= \braket{a | b}
    \end{split}
\end{equation}
which gives the property used in Sec. \ref{sectionOverlaps}.

We can also show that $\ket{G_*}$ must have the property $\bra{a}\overline{\bra{b}} \ket{G_*} = \alpha \braket{a | b}$, $\alpha \in \mathbb{C}$, directly from the boundary condition of Eq. \ref{boundaryCon}. The fact we can express the modes of the boson fields with the modes of the vertex operators means that our boundary condition is also conformal, $L_n\ket{G_*} = \overline{L}_{-n}\ket{G_*}$, where $L_n$ are the Virasoro operators (modes of the stress-energy tensor). Thus, if $\ket{a}$ and $\ket{b}$ are eigenstates of the $L_0$ operator then $\bra{a}\overline{\bra{b}}\ket{G_*}$ can only be non-zero if both $\ket{a}$ and $\ket{b}$ have the same eigenvalue under the application of $L_0$, as $\bra{a}\overline{\bra{b}}L_0\ket{G_*} = \bra{a}\overline{\bra{b}}\overline{L}_0\ket{G_*}$. Next, Let $A$ and $B$ be some generic products of modes of $V_{\pm\epsilon^j}(z)$ (not to be confused of the $A$ and $B$ labels of the two systems in thed main text). From the boundary condition we have $\bra{0}\overline{\bra{0}}A^\dagger \overline{B}^\dagger \ket{G_*} = \bra{0}\overline{\bra{0}}A^\dagger B \ket{G_*} = (\bra{0}A^\dagger B) \overline{\bra{0}}\ket{G_*}$. For this overlap to be non-zero $B^\dagger A \ket{0}$ must have eigenvalue zero under the application of $L_0$. As only the vacuum has eigenvalue zero under the application of $L_0$, $B^\dagger A \ket{0} = (\bra{0}B^\dagger A \ket{0}) \ket{0}$. Hence, $\bra{0}\overline{\bra{0}}A^\dagger \overline{B}^\dagger \ket{G_*} = \bra{0}A^\dagger B \ket{0} \bra{0}\overline{\bra{0}}\ket{G_*}$. As all states are generated by modes of the vertex operators applied on the vacuum we then have in general, $\bra{a}\overline{\bra{b}}\ket{G_*} = \alpha \braket{a | b}$, where $\alpha = \bra{0}\overline{\bra{0}}\ket{G_*}$. We can then see that any state satisfying the boundary condition of Eq. \ref{boundaryCon}, must also satisfy this property, which also implies $\ket{G_*}$ is fixed up to multiplication by a complex number. This was also pointed out by DRR.

\section{Topological Entanglement Entropy} \label{Appendix:Entropy}
As a final check we will now show that our effective description gives the expected topological entanglement entropy. This calculation is somewhat standard now, however we will include it here as some readers may find it useful.

Let us now compute the entropy associated with the partition function given in Appendix \ref{AngularMomentumDiffSection}. Note that as $\alpha, \beta, \gamma \propto 1/\sqrt{N}$, we can treat $\sqrt{N}$ as an effective temperature. Thus, by using Equation \ref{logZ} we find that the entropy in the large $N$ limit is,
\begin{equation}
    \begin{split}
        S =& \sqrt{N} \frac{\partial \ln Z}{\partial \sqrt{N}} + \ln Z \\
        =& \frac{\pi^2}{3} \bigg (\frac{1}{\beta} + \frac{1}{\gamma} \bigg ) - \ln{\text{Vol}(\Gamma)} \\
        &= \text{constant}\sqrt{N} - \ln{\text{Vol}(\Gamma)} \\
    \end{split}
\end{equation}
As $\sqrt{N} \propto L$, where $L$ is the length of the equator of the sphere, we can then identify the topological entanglement entropy $\gamma_{\text{topo}}$ as,

\begin{equation}
    \gamma_{\text{topo}} = \ln{\text{Vol}(\Gamma)}
\end{equation}
One can easily compute the volume of the unit cell of $\Gamma$ to be Vol$(\Gamma) = \sqrt{3}$.

In general the topological entanglement entropy is given by $\gamma_{\text{topo}} = \ln{\sqrt{|\det{K}|}}$, \cite{Kitaev2006, Cano2015, Berger}. From Eq. \ref{KMatrix} we can see that $\det{K} = 3$. Thus, our effective description gives a topological entanglement entropy that agrees with the general result.

\section{Further Plots} \label{Appendix:FurtherPlots}
\begin{figure}[h!]
    \centering
    \includegraphics{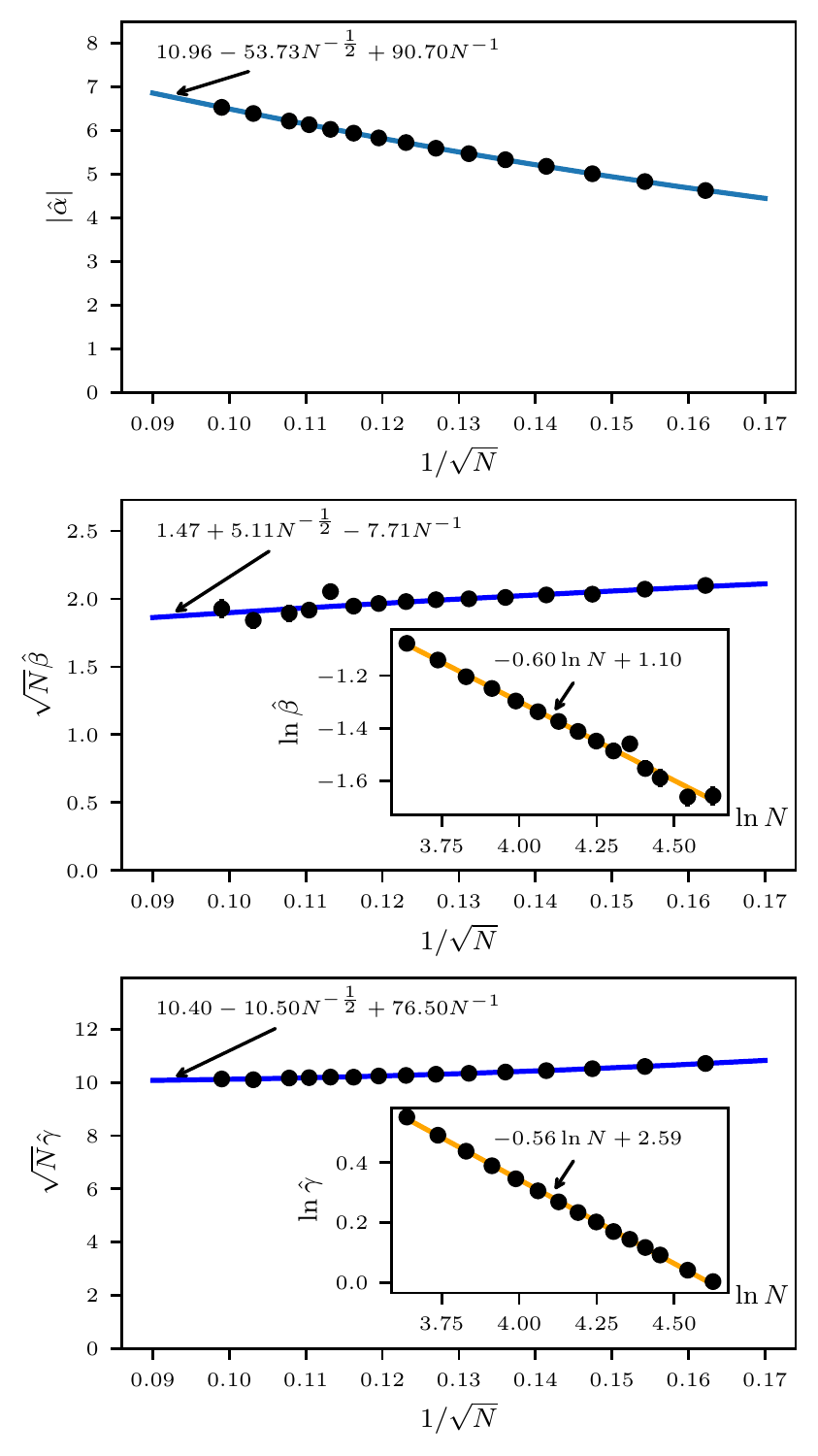}
    \caption{Same as for Fig. \ref{fig:scale2Plot} but for the case of the $S^{\{ 3 \}}_{ES}$ model. The parameters are nearly identical to those of Fig. \ref{fig:scale2Plot} (as one would expect).}
    \label{fig:ScaleTest5ParamAdd}
\end{figure}
Here we show the results of the tests of Sec. \ref{Section:ScaleTest} for the $\hat{\alpha}$, $\hat{\beta}$ and $\hat{\gamma}$ parameters of the $S^{\{ 3 \}}_{ES}$ model. The fitted parameters for this test are nearly identical to those for the $S^{\{ 2 \}}_{ES}$ model. Hence, this data has been given here and not in the main text.

\newpage

\bibliography{references.bib}

\end{document}